\documentclass{aa}  

\usepackage{multirow, graphicx, rotating, array}
\usepackage{natbib}
\usepackage{subfigure, subcaption, caption}
\usepackage{txfonts}
\usepackage{xcolor}
\usepackage{hyperref}
\usepackage{makecell}
\begin{document}

   \title{Studying the variability of the He triplet to understand the detection limits of evaporating exoplanet atmospheres}

   \author{
    Samson J. Mercier\inst{1,2,*},
    Xavier Dumusque\inst{1},
    Vincent Bourrier\inst{1},
    Khaled Al Moulla\inst{1},
    Michael Cretignier\inst{3},
    William Dethier\inst{4},
    Gaspare Lo Curto\inst{5},
    Pedro Figueira\inst{1,4},
    Christophe Lovis\inst{1},
    Francesco Pepe\inst{1},
    Nuno C. Santos\inst{4,6},
    St\'ephane Udry\inst{1},
    Fran\c{c}ois Wildi\inst{1},
    Romain Allart\inst{7},
    Fr\'ed\'erique Baron\inst{7,8},
    Fran\c{c}ois Bouchy\inst{1},
    Andres Carmona\inst{9},
    Marion Cointepas\inst{1,9},
    Ren\'e Doyon\inst{7,8},
    Yolanda Frensch\inst{1,5,},
    Nolan Grieves\inst{1},
    Lucile Mignon\inst{1,9},
    Louise D. Nielsen\inst{1,10,11}
    }

   \institute{
    \inst{1}Observatoire de Gen\`eve, D\'epartement d’Astronomie, Universit\'e de Gen\`eve, Chemin Pegasi 51, 1290 Versoix, Switzerland\\
    \inst{2}Department of Earth, Atmospheric and Planetary Sciences, Massachusetts Institute of Technology, Cambridge, MA 02139, USA\\
    \inst{3}Department of Physics, University of Oxford, Oxford OX13RH, UK\\
    \inst{4}Instituto de Astrof\'isica e Ci\^encias do Espa\c{c}o, Universidade do Porto, CAUP, Rua das Estrelas, 4150-762 Porto, Portugal\\
    \inst{5}European Southern Observatory (ESO), Av. Alonso de Cordova 3107,  Casilla 19001, Santiago de Chile, Chile\\
    \inst{6}Departamento de F\'isica e Astronomia, Faculdade de Ci\^encias, Universidade do Porto, Rua do Campo Alegre, 4169-007 Porto, Portugal\\
    \inst{7}Institut Trottier de recherche sur les exoplan\`etes, D\'epartement de Physique, Universit\'e de Montr\'eal, Montr\'eal, Qu\'ebec, Canada\\
    \inst{8}Observatoire du Mont-M\'egantic, Qu\'ebec, Canada\\
    \inst{9}Univ. Grenoble Alpes, CNRS, IPAG, F-38000 Grenoble, France\\
    \inst{10}European Southern Observatory (ESO), Karl-Schwarzschild-Str. 2, 85748 Garching bei M\"unchen, Germany\\
    \inst{11}University Observatory, Faculty of Physics, Ludwig-Maximilians-Universit\"at M\"unchen, Scheinerstr. 1, 81679 Munich, Germany\\
    \inst{*}\email{merci228@mit.edu}
    }

   \date{Received 04 November 2024 / Accepted 15 January 2025}

  \abstract
    {With more than a dozen significant detections, the helium triplet has emerged as a key tracer of evaporating exoplanet atmospheres. This near-infrared feature can be observed from the ground and holds great promise, especially with upcoming observations provided by new-generation instruments such as the Near Infrared Planet Searcher (NIRPS). However, as the helium triplet is also present in stellar spectra, careful removal of the average stellar contribution is necessary to accurately characterize the atmospheres of transiting exoplanets. In this study, we analyze multi-epoch observations of the Sun obtained with NIRPS to investigate the temporal variability of the helium triplet. Our findings reveal significant variability across different timescales, ranging from minutes to days. We identify telluric contamination and stellar activity as likely sources for the short-term and long-term variability, respectively. Importantly, we demonstrate that this variability has minimal impact on the retrieval of planetary parameters crucial to the study of atmospheric escape.}


   \titlerunning{Studying the variations of the He triplet}
   \authorrunning{S.J. Mercier et al.}
   \maketitle
%

\nolinenumbers

\section{Introduction}\label{sec:Introduction}
Since the discovery of the first exoplanet around a solar-like star, 51 Peg b \citep{Mayor1995}, the development of increasingly precise and optimized instruments to measure planetary mass (e.g., HARPS\footnote{High Accuracy Radial velocity Planet Searcher}, \citealt{Mayor2003}; HARPS-N\footnote{High Accuracy Radial velocity Planet Searcher for the Northern hemisphere}, \citealt{Cosentino-2012}; ESPRESSO\footnote{Echelle SPectrograph for Rocky Exoplanet and Stable Spectroscopic Observations}, \citealt{Pepe2021}; EXPRES\footnote{EXtreme PREcision Spectrometer}, \citealt{Jurgenson2016}; NEID\footnote{NN-explore Exoplanet Investigations with Doppler spectroscopy}, \citealt{Schwab2016}) and radius (e.g., CoRoT\footnote{Convection, Rotation and planetary Transits}, \citealt{Auvergne2009}; Kepler, \citealt{Borucki2010}; TESS\footnote{Transiting Exoplanet Survey Satellite}, \citealt{Ricker2015}) has facilitated the detection and characterization of a growing number of exoplanets. 
In particular, studying their atmospheres provides insight into the physical properties of these planets, namely their chemical composition and surface temperature. Hot exoplanets, such as hot Jupiters, are ideal candidates for atmospheric studies given that their H and He-dominated atmospheres are significantly extended and have large scale heights. The proximity of these planets to their host stars results in intense heating of their thermobase by stellar radiation, causing hydrodynamic expansion and, in some cases, loss of gas from the upper atmospheric layers \citep{Lammer2003, Vidal-Madjar2003}. 
Atmospheric mass loss not only provides insights into the physical properties of exoplanets but can also help explain some of the puzzling observations seen in population studies. For instance, the lack of short-period Neptune-mass planets, also known as the hot Neptune desert, could be due to the fact that some planets are not massive enough to retain their expanding gaseous atmosphere \citep[e.g.,][]{Lecavelier2007, Davis2009, Beauge2013, Lundkvist2016, Mazeh2016, Owen2019}. Additionally, the clear gap in the radius distribution between $1.5 \ R_{\oplus}$ and $2 \ R_{\oplus}$, otherwise known as the radius valley, may be caused by some exoplanets retaining a substantial atmosphere while others losing it through photoevaporation \citep[e.g.,][]{Lopez2012, Lopez2013, Owen2013, Owen2017, Fulton2017, Rogers2021} or core-powered mass loss \citep[e.g.,][]{Ginzburg2016, Gupta2019, Gupta2020, Gupta2021}. 
Therefore, developing techniques to study the impact of various atmospheric escape mechanisms is crucial to improving our understanding of the evolution and demographics of exoplanets. 

Historically, studying the H Ly-$\alpha$ line through transmission spectroscopy has been a prominent method to explore atmospheric escape. It has yielded significant results, especially for hot Jupiters orbiting bright stars like HD 209458b \citep{Vidal-Madjar2003, Vidal-Madjar2008, Ehrenreich2008} and HD 189733b \citep{Lecavelier2010, Lecavelier2012, Bourrier2013, Bourrier2020}. However, as H Ly-$\alpha$ is in the ultraviolet range, it is not observable from the ground and space-based observations are plagued with interstellar medium (ISM) absorption and geocoronal emission \citep{Landsman1993, Wood2002, Wood2005}. Although these difficulties can be overcome by restricting the spectral analysis to the wings of H Ly-$\alpha$, ISM absorption will in any case limit observations to extrasolar systems in close proximity to the Sun. 

Another possible tracer of escaping atmospheres, initially theorized by \citet{Seager2000}, is the metastable HeI (2$^3$S) triplet. This transition presents features at 10832.1, 10833.2, and 10833.3$\ \mathrm{\AA}$, a wavelength region devoid of ISM and geocoronal absorption \citep{Indriolo2009}. Furthermore, as these lines are in the near-infrared (IR) they can be studied with dedicated high-resolution ground-based instruments, such as HPF\footnote{Habitable-zone Planet Finder} \citep{Mahadevan2012, Mahadevan2014}, SPIRou\footnote{SPectropolarimètre InfraROUge} \citep{Donati2020}, CARMENES\footnote{Calar Alto high-Resolution search for M dwarfs with Exoearths with Near-infrared and optical Échelle Spectrographs} \citep{Quirrenbach2014}, NIRPS \citep{Bouchy2017}, or NIGHT\footnote{Near-Infrared Gatherer of Helium Transits}, which is currently in development \citep{Jentink2024}.
The He triplet has successfully been detected in several exoplanets, including, but not limited to, WASP-107b \citep{Spake2018, Allart2019, Kirk2020, Spake2021},  HAT-P-11b \citep{Allart2018, Mansfield2018}, HD 189733b \citep{Salz2018, Guilluy2020}, and most recently WASP-69b \citep{Nortmann2018, Vissapragada2020, Vissapragada2022}. It is known that the He triplet's core is formed at the chromosphere level and therefore is sensitive to magnetic fields \citep{Guilluy2020}. Magnetic fields are not supposed to evolve significantly over the timescale of a transit and it is therefore assumed that the He triplet is stable over such a timescale. However, the He triplet could also be sensitive to other stellar noises such as oscillations, granulation, and super-granulation that evolve on timescales of a few minutes to a few days, or be contaminated by telluric lines that can also evolve significantly over a few hours. In this paper, we investigate the variability of the solar He triplet over a few hours for several days and how the observed variability impacts the signature of an exoplanet atmosphere. Additionally, we search for links between this variability and stellar and telluric contamination tracers to better understand the origin of this helium transition's variations. To this end, the He triplet's wavelength range offers a unique advantage as it includes both a telluric water line and a stable photospheric SiI line. This positioning allows us to effectively separate telluric and stellar contributions and study them independently.

We provide a brief outline of the NIRPS instrument and the observations used in Section \ref{sec:NIRPS}. We describe the methods used to study variations in the He triplet properties, their possible causes, and their effects on retrieved planetary parameters in Sect. \ref{sec:Methods}. Finally, we present and discuss the implications of our results in Sect. \ref{sec:results_and_discussion}. 

\section{NIRPS observations}\label{sec:NIRPS}
NIRPS is a high-resolution, near-IR spectrograph installed on the 3.6m telescope at the European Southern Observatory's (ESO) La Silla site in Chile, along with the HARPS spectrograph. One of the goals of this instrument is to study the atmospheres and orbital architectures of a variety of exoplanets using high-resolution time-series spectroscopy, in both transmission and emission \citep[][Bouchy et al. 2024 in prep.]{Artigau:2024aa}. Such studies could improve our understanding of the mechanisms driving exoplanet formation and evolution, as well as observed trends in exoplanet demographics. NIRPS complements the spectral range of HARPS, by covering the Y, J, and H bands ($0.97 - 1.81 \ \mu m$), and offers two modes of observation depending on the magnitude of the observed star: a high-accuracy mode (HA) and a high-efficiency mode (HE) that can reach spectral resolutions of 90,000 and 75,000, respectively. The HE mode, which uses a larger science fiber, is less affected by modal noise and thus provides more precise retrievals of atomic and molecular features from atmospheric spectra. We therefore only analyze the HE mode solar data in this paper.

The HARPS experiment for light integrated over the Sun (HELIOS) is installed outside the 3.6m telescope building. It consists of a small robotic telescope that focuses the disk-integrated Sun's light onto an integrating sphere and feeds it to HARPS and NIRPS via optical fibers. This telescope therefore allows for the acquisition of high-precision solar spectra for several hours per day, on any possible day. As we aim to study the variability of the He triplet over different timescales, we extracted the NIRPS spectra for multiple days of HELIOS observations. Details concerning these observations can be found in Table \ref{tab:Observation_table}. 

Further insights into the design and performance of NIRPS and HELIOS are provided by \cite{Wildi2017, Blind2022, Cointepas2022, Frensch2022, Thibault2022, Wildi2022, AlMoulla2023, Bouchy2024} and the dedicated ESO webpages\footnote{\url{https://www.eso.org/sci/facilities/develop/instruments/NIRPS.html}}$^{,}$\footnote{\url{https://www.eso.org/public/announcements/ann18033/}}.

\section{Methodology}\label{sec:Methods}

The following sections describe the methods used to characterize the He triplet variability and study its impact on retrieved planetary parameters. As multiple analyses are done on different portions of the data, we clarify what is done on what in the following points: 
\begin{enumerate}
   \item The long-term variability analysis (Sects. \ref{sec:long_timescale} and \ref{sec:discussion_long_variability}, and Fig. \ref{fig:Long_timescales_+_RHK}) was carried out with all available observation days.
    \item The short-term variability analysis (Sects. \ref{sec:diagnostics} and \ref{sec:discussion_short_variability}, Appendix \ref{sec:He-Si_Correlation_Normalization} and \href{https://zenodo.org/records/14674319}{E}, and Figs. \ref{fig:master_and_residuals}, \ref{fig:best_fit_He_param} and \ref{fig:best_fit_He_param_27}) was carried out with all days of observation. All days presented similar results, but we opted to present only the results for January 26, 2023 and January 27, 2023 as these two provided the most adequate representation of the trends witnessed across all days.
    \item The injection-recovery tests (Sect. \ref{sec:planet_signal} and \ref{sec:discussion_atm_retrieval}, Appendix \ref{sec:appendix_plan_signal_simul}, and Figs. \ref{fig:mock_abs_spectrum}, \ref{fig:planet_simulation}, and \ref{fig:big_planet_simulation}) were carried out with all days of observation. We present the results for January 26, 2023 and report the results of all other analyses in Table \ref{tab:simulation_results_table}.
\end{enumerate}

\subsection{Data extraction and reduction}\label{sec:reduction}
The raw NIRPS data, stored in the FITS format,  were first processed with the latest version of the standard NIRPS pipeline\footnote{Details can be found in the NIRPS user manual at \url{https://www.eso.org/sci/facilities/lasilla/instruments/NIRPS/doc/manuals/espdr-pipeline-manual-2.4.16.pdf}}. Primary data products were then extracted, which included the wavelength solution in vacuum, raw and telluric-corrected spectra, and properties of the cross-correlation function (CCF) such as its radial velocity (RV), full-width at half-maximum (FWHM), and bisector span. We note that the wavelength solution for the raw spectra and model telluric spectra, used to perform the telluric correction, were the same and were both expressed in the stellar rest frame. Subsequent steps, outlined in the sections below, focused on reducing the data to remove effects that degrade the spectra and have the potential to influence our analyses. The following reduction and analyses focused solely on data from Échelle order 14 (physical order 135) that contains the He triplet.

\subsubsection{Telluric correction}\label{sec:telluric}
Ground-based observations are prone to contamination from absorption and emission lines of various atmospheric species in Earth's atmosphere. These features that can bias stellar spectra and subsequent RV measurements must be removed. While at low resolution individual telluric lines cannot be resolved and removed, this becomes possible when moving to high resolution.

\begin{figure}
    \centering
    \includegraphics[width = \columnwidth]{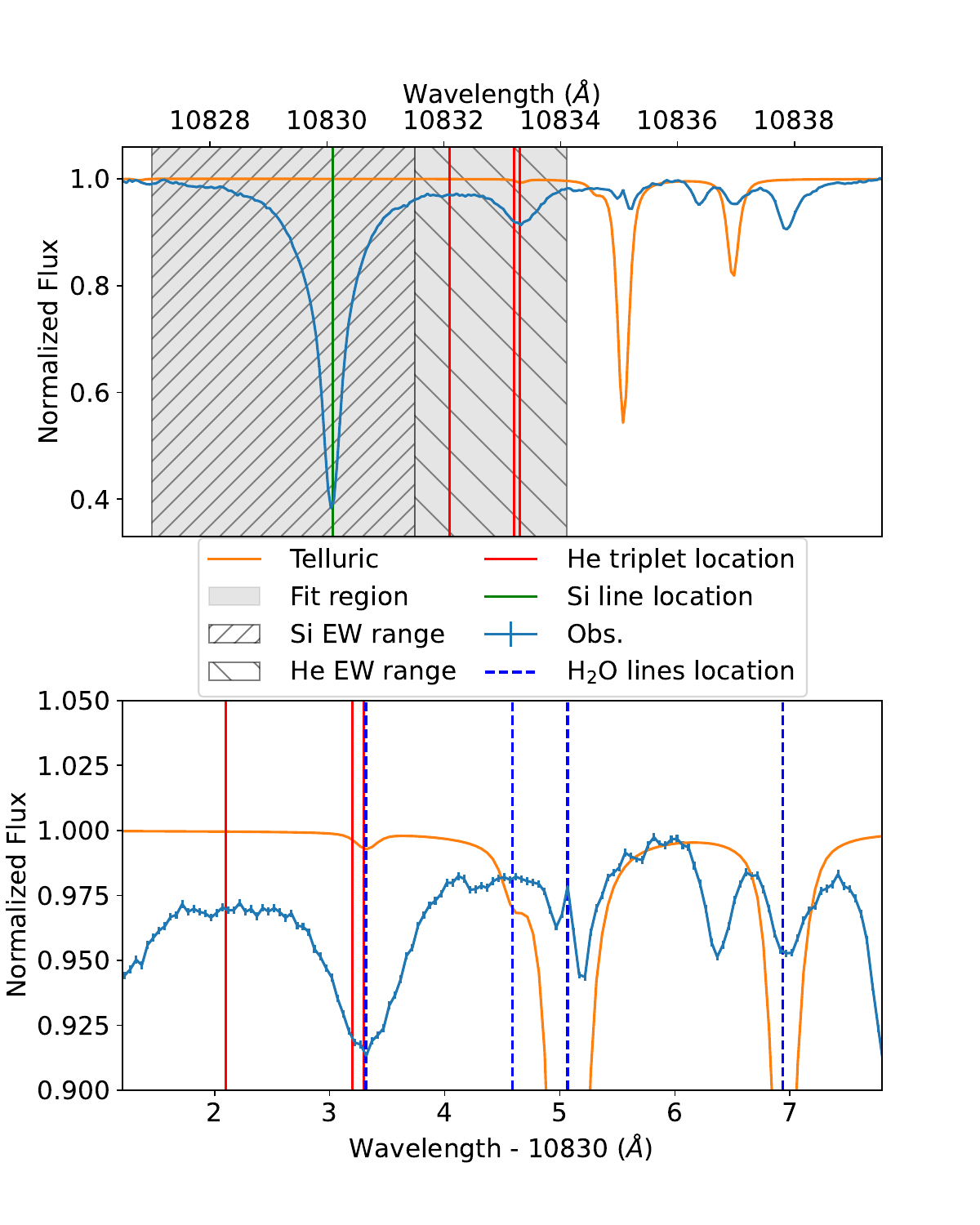}
    \caption{Telluric spectrum for the observation day January 26, 2023. Top panel: The telluric spectrum (orange) used to correct the solar spectrum (blue). The green vertical line marks the position of the SiI line near the He triplet, while the gray band represents the region over which we fit the He triplet to extract its properties (see Sect. \ref{sec:model_fit_Si_He} and Sect. \ref{sec:model_construction}). The diagonal and anti-diagonal hashed regions indicate the range used to calculate the equivalent width (hereafter referred to as EW) of the SiI line and He triplet, respectively. Bottom panel: Zoomed-in view of the region around the He triplet. The position of telluric H$_2$O lines, obtained from the HITRAN database, are indicated by blue dashed vertical lines \citep{HITRAN2022}. Notably, an H$_2$O absorption line overlaps with the He triplet around $10833.3 \ \AA$. The impact of this line on the He triplet is explored in Sect. \ref{sec:diagnostics} and discussed in Sect. \ref{sec:discussion_short_variability}. In both plots, red vertical lines denote the positions of the He triplet transitions.}
    \label{fig:telluric_spctr}
\end{figure}

In the NIRPS YJH bands, molecules such as H$_2$O, O$_2$, CO$_2$, and CH$_4$ have the most significant impact. The absorption linked to these molecules can be modeled in first order from the airmass, except for water vapor, which requires knowledge of the integrated water column density and its variations linked to the weather conditions.
The telluric correction used is included in the NIRPS processing pipeline and follows the methodology presented by \cite{Allart2022}. It involves building a model transmission spectrum of Earth's atmosphere from a line-by-line radiative transfer code assuming a single atmospheric layer. The spectral lines of modeled species are retrieved from the HITRAN database\footnote{\url{https://hitran.org}}. For each telluric species, a Levenberg-Marquardt algorithm minimizes the $\chi^2$ between the CCF derived from the model using a few strong telluric lines of this species and the CCF derived from the observation using the same line selection. The resulting best-fit is used to remove the telluric lines. \cite{Allart2022} demonstrate a correction better than 2\% peak to valley across spectral types F to K dwarfs with this pipeline. Although their tests are performed on visible spectra, the near-IR region around the He triplet is not significantly contaminated by tellurics and thus we expected a similar correction level. An example best-fit telluric spectrum is shown in Fig. \ref{fig:telluric_spctr}.

\subsubsection{Blaze, dispersion, and black body correction}\label{sec:DLL}
We corrected each raw solar spectrum for the Blaze function using the corresponding Tungsten-lamp calibration spectrum. 
Following this correction, the flux of each spectrum presented a slow decreasing monotonous trend. Although the trend could have been ignored as it was negligible over the He triplet's wavelength range, we removed it to avoid any bias that could influence later diagnostics or analyses. The trend arose from two sources, thus we removed both their contributions to ensure the robustness of our correction.
We first accounted for the wavelength-dependent dispersion of the spectrograph (i.e., the fact that the width in wavelength of a pixel, and therefore the amount of light it collects, varies as a function of wavelength) as it had the strongest contribution to the decreasing monotonous trend. We used the DLLDX product, computed by the standard NIRPS pipeline and recorded in the FITS files, to detrend our spectra from this effect. This product records the size of each pixel in the wavelength domain.
Secondly, because we are probing the Sun's spectral energy distribution (SED) in the IR, we had to account for the shape of the Planck function over the wavelength range considered. We modeled the Sun's SED at an effective temperature $T_{\mathrm{eff}} = 5778 $ K with Astropy's BlackBody function and detrended our spectra with this model.

\subsubsection{S/N, RV, and airmass clipping}\label{sec:outlier_clipping}
We excluded spectra affected by poor weather, telluric-, instrumental-, or stellar-induced anomalies by applying sigma-clipping to the signal-to-noise ratio (S/N) and RV values. Spectra deviating from the daily average by more than $20 \sigma$ were discarded. This threshold was selected because it was the lowest value that effectively removed outliers from all datasets.
Although S/N and RV outliers often overlap, we applied both steps for thoroughness. Additionally, we omitted data from observations taken at airmass greater than two that can present additional telluric contamination.

\subsubsection{Normalization and interpolation}\label{sec:normalization}
We performed normalization to remove variations in the continuum level across different spectra, thus simplifying the subsequent steps of our analysis. To do so, we applied the RASSINE framework, outlined by \citet{Cretignier2020}. Notably, we forced RASSINE's alpha-shape algorithm to avoid the narrow region around the He triplet, thus guaranteeing that none of the variations intrinsic to the He triplet were absorbed by the normalization.

To accurately study the properties of the He triplet across different timescales, all spectra had to be interpolated on a common wavelength grid. RASSINE automatically performs this interpolation. Therefore, by normalizing the spectral time series for each day with RASSINE, we simultaneously interpolated all spectra onto a common wavelength solution.

\subsection{He triplet variability analysis}\label{sec:variability}
Our analysis of variability aims at characterizing temporal changes in the helium triplet and elucidating the underlying causes. In the following sections, we highlight the different strategies employed to pursue this goal. 

In particular, we conducted a preliminary assessment of the He triplet variability by comparing individual spectra to a master spectrum and visually inspecting the residuals in search of variations across the day of observation. Subsequently, we performed a rigorous analysis of the variability by fitting the He triplet line with a custom model and studying the variations in the fitted properties. Finally, we describe the different proxies that were used to investigate the origin of the He triplet variability.

\subsubsection{Master spectrum}\label{sec:master}
\begin{figure}[h!]
    \centering
    \includegraphics[width = \columnwidth]{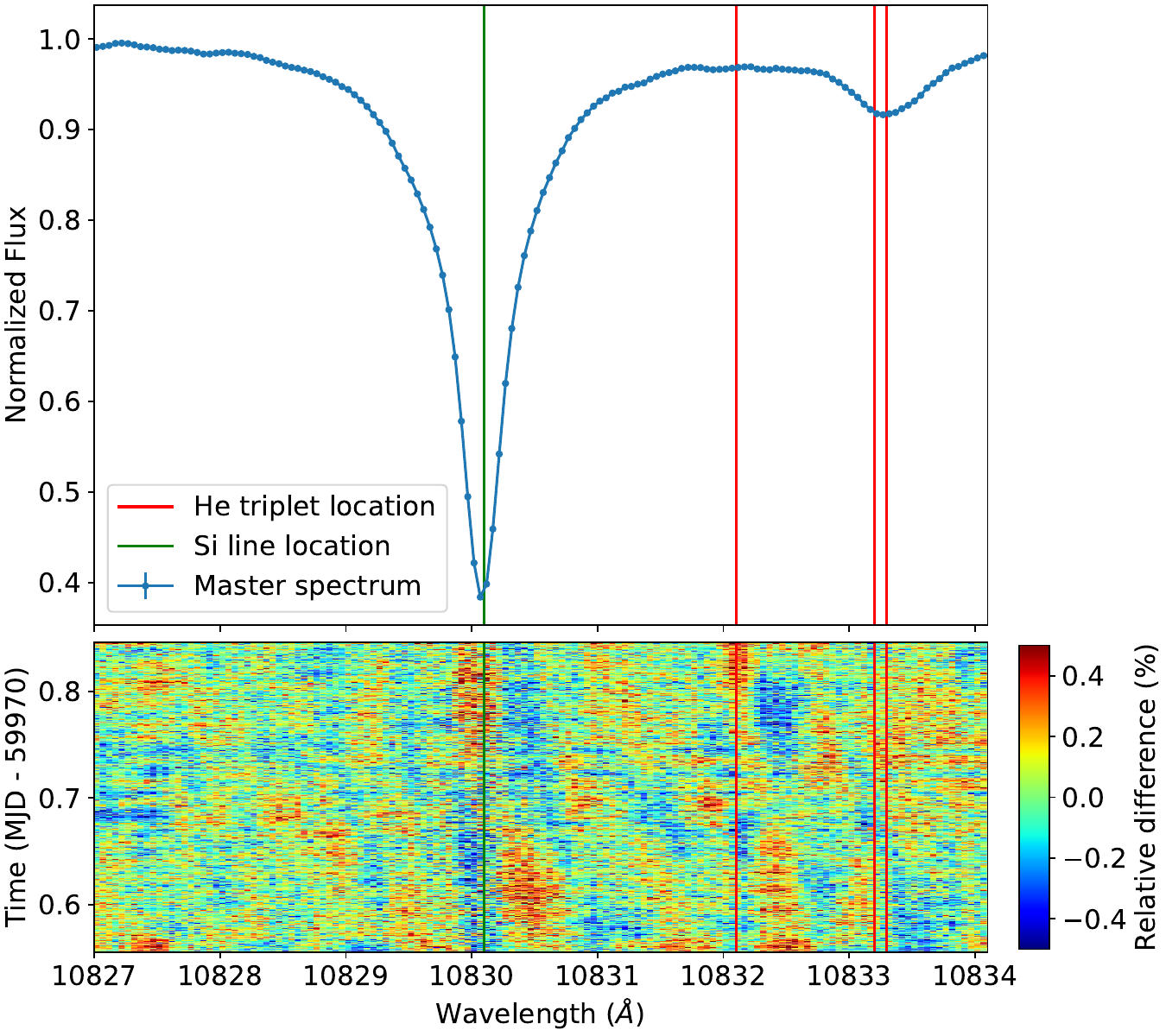}
    \caption{Master spectrum and corresponding residual map for the observation day January 26, 2023. Top panel: Master spectrum retrieved from the RASSINE pipeline, as outlined in Sect. \ref{sec:master}. Bottom panel: Residual map, in the solar rest frame, obtained by calculating the relative difference between the master spectrum and each individual spectrum of that day. The positions of the Si line and He triplet lines are marked with green and red vertical lines, respectively.}
    \label{fig:master_and_residuals}
\end{figure}

The master spectrum is a direct output of the RASSINE pipeline, built by summing all the raw spectra from the provided spectral time series and subsequently normalizing this sum. This normalization uses the same anchor point locations as those used to normalize the spectral time series (see Sect. \ref{sec:normalization}).
By comparing each normalized spectrum to the master we built a residual map, as shown in Fig. \ref{fig:master_and_residuals}. This map presents structures around the He triplet and SiI lines that we analyzed to determine their origin. Shifting the maps from the stellar reference frame to the Earth's barycentric frame showed no transformation of the diagonal tracks, seen around 10830 and 10832 $\AA$, into vertical features, ruling out telluric or instrumental contamination. This indicated that the features had to be of solar origin. Additionally, residual maps from other observation days exhibited radically different structures, further supporting the conclusion that these features were driven by solar processes.

\subsubsection{Radiative transfer}\label{sec:model_fit_Si_He}
We obtained a more detailed characterization of the He triplet properties and their variations by modeling the He absorption in the solar chromosphere with a plane-parallel single-layer approximation. The equations used in this routine, which we describe below, are based on previous works aimed at studying the impact of radiation pressure on interstellar gas \citep[e.g.,][]{Lagrange1998}.

To model the observed line profiles of the solar spectrum ($F_{\rm obs}$), we used the solution of the radiative transfer equation considering only extinction and ignoring the three-dimensional structure of the solar atmosphere.
\begin{equation}\label{eq:rad_transfer_theory}
    F_{\rm obs} = F_0\times e^{-\tau_0}
\end{equation}
Here, $\tau_0$ is the solar atmosphere's total optical depth, and $F_0$ is the base profile, taken as the spectrum at the base of the solar atmosphere. Given the scope of this project, we considered this approach to be a reasonable compromise between accuracy and computational efficiency. We considered a limited number of atomic transitions in the absorption process, namely, the He triplet transitions and the neighboring SiI transition. Including the SiI line ensured an accurate retrieval of the He triplet properties given that its red tail blends with the He triplet feature around 10832.1 $\AA$. As a result, the absorption, captured in the $e^{-\tau_0}$ component and that is applied to the base profile $F_0$ provided as input to our model, is calculated as follows:
 
Since all transitions are considered simultaneously $\tau_0 = n_{\mathrm{He}}\times(\sigma_{1, \mathrm{He}} + \sigma_{2, \mathrm{He}} + \sigma_{3, \mathrm{He}}) + n_{\mathrm{Si}}\times\sigma_{\mathrm{Si}}$, 
with $n_\mathrm{He}$, $n_\mathrm{Si}$ the column densities of He and Si, and $\sigma_{1-3, \mathrm{He}}$, $\sigma_{\mathrm{Si}}$ the cross-sections of the transitions considered. 
Each cross-section, exemplified by $\sigma_0$, is determined as follows:
 \begin{equation}
     \sigma_0(\lambda) = \frac{e^2}{4 \epsilon_0 \times m_e \times c \times \sqrt{\pi}} \times \frac{f_{\rm osc} \times \lambda_0 \times L(\lambda)}{w}
 \end{equation}
The left-hand side contains fundamental constants with $e$ the elementary charge, $m_e$ the electron mass, $\epsilon_0$ the vacuum permittivity, and $c$ the speed of light. The right-hand side consists of terms depending on intrinsic properties of the electronic transition considered with $f_{\rm osc}$ the oscillator strength, $\lambda_0$ the reference wavelength in vacuum, $L$ the line profile, and $w$ the line width. While the $1/w \sqrt{\pi}$ factor is due to the normalization of the line profile, the $\lambda_0$ factor arises from approximating the cross-section as a Dirac delta function centered on the reference wavelength.
Two line profiles are implemented in our model. A Gaussian profile,
\begin{equation}\label{eq:gauss_line_profile}
    L(\lambda) = \mathrm{exp} \left({-\frac{1}{w^2}\left(\frac{c}{\lambda_{0}} (\lambda - \lambda_{0}) + rv \right)^2}\right),
\end{equation}
and a Voigt profile, 
\begin{equation}\label{eq:voigt_line_profile}
\begin{aligned}
    z &= \frac{1}{w}\left(\frac{c}{\lambda_{0}} (\lambda - \lambda_{0}) + rv \right) + i \gamma  \\
    L(\lambda) &= \mathrm{Re}[W(z)].
\end{aligned}
\end{equation}
In Eq. \ref{eq:voigt_line_profile}, $\gamma$ is the Voigt damping parameter and $W$ is the Faddeeva function.
In Eqs. \ref{eq:gauss_line_profile} and \ref{eq:voigt_line_profile}, $\lambda$ is the wavelength grid assigned to the input profile $F(\tau_0)$ and $rv$ is a radial velocity offset. The line width, $w$, is defined as follows:
\begin{equation}\label{eq:width}
    w^2 = \underbrace{\frac{2\times k_B \times T}{m}}_{\substack{\text{Thermal} \\ \text{broadening}}} + \underbrace{\left(\frac{c}{R}\right)^2}_{\substack{\text{Instrumental} \\ \text{broadening}}} + \underbrace{(v \cdot \mathrm{sin}(i))^2}_{\substack{\text{Rotational} \\ \text{broadening}}}
\end{equation}
Here, $k_B$ is the Stefan-Boltzmann constant, $T$ and $m$ are the temperature and mass of the species considered, $i$ and $v$ are the stellar inclination and rotational velocity, and $R$ is the instrumental resolving power. All variables are expressed in the International System of Units (SI). The equations for the helium and silicon transitions are identical to Eqs. \ref{eq:gauss_line_profile}, \ref{eq:voigt_line_profile}, and \ref{eq:width}, except for the values of $f_{\rm osc}$, $\lambda_{0}$, and $m$ that were replaced by those specific to each species and transition. All the physical constants were fixed to values according to the SI. $v\mathrm{sin}(i)$ was set to the Sun's projected rotational velocity \citep[1.97 km/s, ][]{Lang2013}, and $R$ was set to the average resolving power of the instrumental mode used ($R_{\rm HE} = 74,000$). The values for $f_{\rm osc}$ and $\lambda_0$ were extracted from the NIST Atomic Spectral Database\footnote{\url{https://www.nist.gov/pml/atomic-spectra-database}}.

By selecting an appropriate base profile for the solar spectra, such as the continuum, we fitted for these variables and retrieved the He triplet properties. Since we were fitting normalized solar spectra, the continuum profile was described with a constant flux level whose value was adjusted to each fitted spectrum.

\subsubsection{Model construction and fitting}\label{sec:model_construction}
Using the model detailed in Sect. \ref{sec:model_fit_Si_He}, we fitted the solar spectra to retrieve the He triplet properties. To ensure an accurate retrieval, we included the SiI line at 10830 $\AA$ in the model. Indeed, this line's red tail blends with the He triplet and could otherwise have biased the fit.

The SiI line exhibited a complex line shape that neither a Gaussian nor Voigt profile can capture individually. Thus, we adopted a hybrid approach, fitting its wings with a Voigt profile and its core with a Gaussian profile. In contrast, each He triplet line was well-described with a Gaussian profile. This setup led us to choose a fitting range, fixed for all spectra, that included the He triplet and SiI lines as much as possible while excluding other neighboring lines. The selected fit range is illustrated in Fig. \ref{fig:telluric_spctr}.

The final model included eight free parameters: shared temperature, column density, and radial velocity offset for all three He triplet transitions, separate values for the SiI transition, a damping parameter for the SiI wings' Voigt profile, and a multiplicative factor controlling the continuum level.
We fitted each telluric-corrected, normalized solar spectrum using Scipy's curve\_fit function. As bounds were set for all free parameters, the Trust Region Reflective algorithm was needed to perform the minimization step of the nonlinear least squares method. 

Given our hybrid approach to modeling the SiI line, we needed to determine the wavelength borders that defined the core and wings of the line. To do so, we first computed the Gaussian and Voigt profiles for the SiI line with the set of parameters at the minimization step considered. Using a least-square approach, we then found the wavelengths on either side of the SiI line core that minimized the overlap between the Gaussian and Voigt profiles. These wavelengths were used to compute a final model describing the full profile of the SiI line. We note that minimizing the overlap between the core and wings was crucial to creating a model that was as continuous as possible across the full SiI line.

After fitting, He triplet and SiI EWs were calculated for each normalized solar spectrum by integrating the flux within the wavelength ranges of the respective lines. The wavelength range was computed from the absolute transition wavelength (shown in Fig. \ref{fig:telluric_spctr}) corrected from the fitted radial velocity offset of the He triplet and SiI lines. The continuum level used to perform this integration was set to the best-fit value of the multiplicative factor controlling the continuum profile. 

\begin{figure}
    \centering
    \includegraphics[width=\columnwidth]{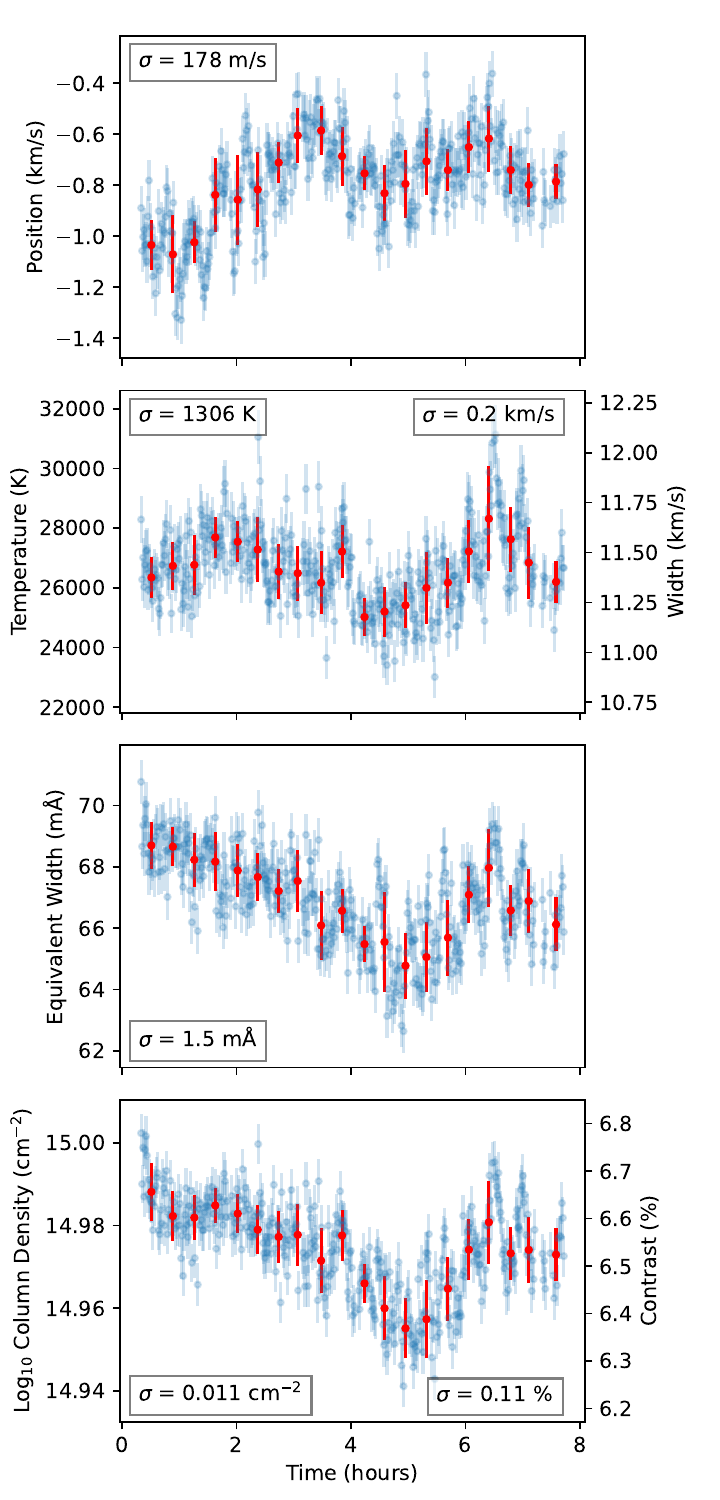}
    \caption{Time series of He triplet properties for the observation day January 26, 2023. From top to bottom: He triplet position,  width, EW, and depth. The depth and width are expressed in terms of temperature and column density, the units used in our custom model, as well as km/s and \%, for better comprehension. The full time series is shown in blue points, and 20$\times$ ($\sim$ 19 min) bins are shown as red points. The standard deviation of each time series is indicated in their respective legends.}
    \label{fig:best_fit_He_param}
\end{figure}

\begin{figure}
    \centering
    \includegraphics[width=\columnwidth]{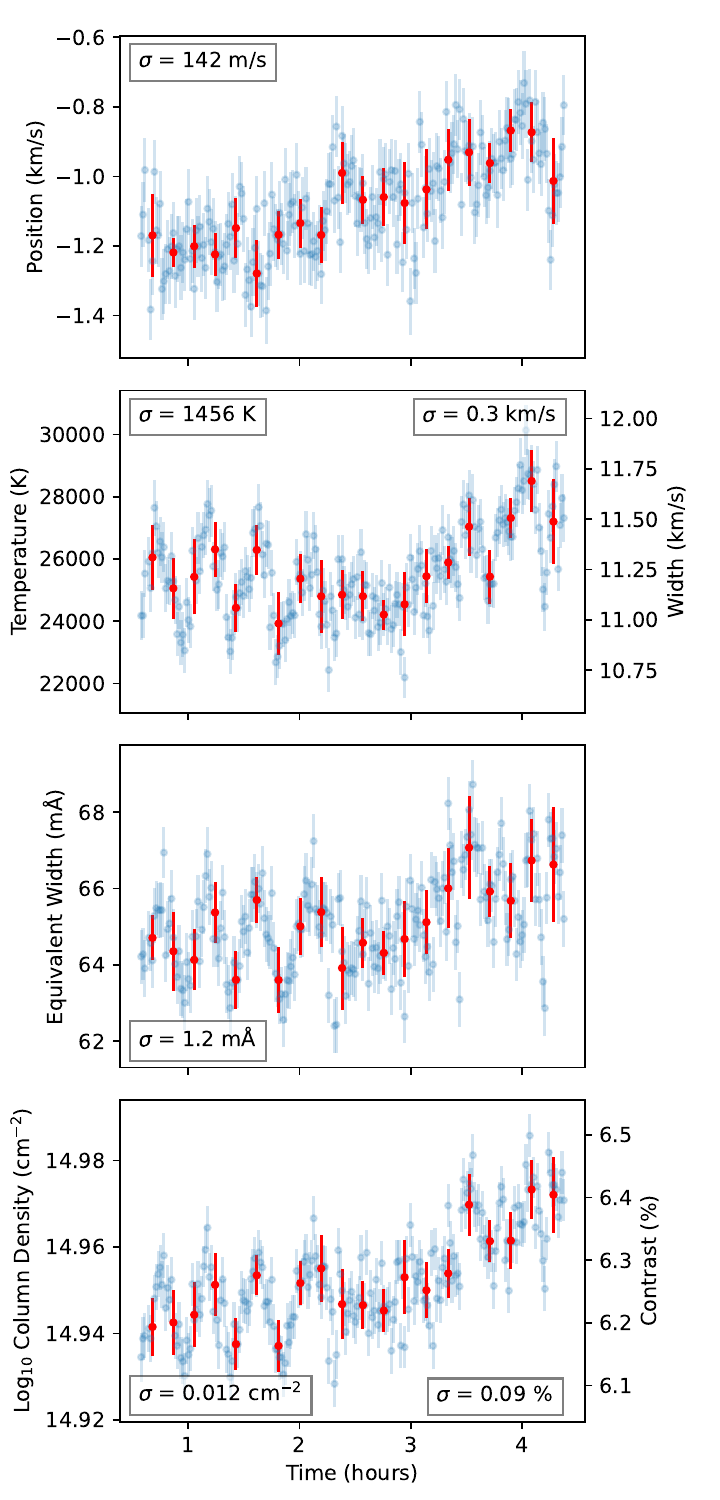}
    \caption{Same as Fig. \ref{fig:best_fit_He_param}, except for the observation day January 27, 2023.}
    \label{fig:best_fit_He_param_27}
\end{figure}

Additionally, we improved our estimate of the best-fit parameter errors using a bootstrap approach. To do so, we generated k=500 realizations of each solar spectrum by sampling a Gaussian distribution with mean and standard deviation corresponding to the value and error on the flux of each point in the spectrum. The uncertainty on each original best-fit parameter was approximated by its corresponding standard deviation across the best-fits of the k realizations.
Finally, for a more tangible measure of the  He triplet's width and depth, we converted the best-fit He triplet temperatures and column densities to line widths in km/s and contrasts in \%. The line widths were obtained by evaluating Eq. \ref{eq:width} with the best-fit temperatures and retrieving $w$. To determine the contrast for each solar spectrum, we subtracted the flux at the center of the model He triplet from the continuum flux level, retrieved from the fitted multiplicative factor.
We studied the evolution of the best-fit depth, position, width, and EW for the He triplet by plotting their time series as illustrated in Figs. \ref{fig:best_fit_He_param} and \ref{fig:best_fit_He_param_27}. These measurements are discussed in detail in Sect. \ref{sec:discussion_short_variability}.


\subsubsection{Variability diagnostics}\label{sec:diagnostics}
Changes in the He triplet properties measured in our spectra could potentially be of telluric and/or stellar origin. In an effort to identify the true underlying cause, we examined proxies associated with each of these mechanisms.

Telluric contamination due to imperfect correction could arise from temporal variations in atmospheric components. In particular, the presence of a telluric water line overlapping with the He triplet prompted us to explore this possibility (see Fig. \ref{fig:telluric_spctr}). These variations can be assessed by studying the total amount of a particular species present in the vertical column of air above the telescope, also known as the integrated column density (ICD). We retrieved the ICD for H$_2$O, O$_2$, and CO$_2$ from the NIRPS pipeline FITS files. As expected, the time series of these quantities were strongly correlated with the airmass. Thus, to study the unbiased effects of ICDs we detrended them using the airmass measurements provided in the NIRPS pipeline FITS files. These same measurements were used to explore the impact of airmass on the He triplet. In addition, we also fitted the water line overlapping with the He triplet with the following Gaussian profile to extract and subsequently study its properties. 
\begin{equation}\label{eq:gaussian_linear}
   G = \left(1 - a \times \mathrm{exp} \left(-\frac{1}{2} (\frac{x-\mu}{\sigma})^2\right)\right) \times (b \times x+c),
\end{equation}
where $a$, $\mu$, and $\sigma$ are the depth, position, and width of the line, while $b$ and $c$ are the first-order polynomial coefficients describing the local continuum. The fitting procedure was performed with the curve\_fit function. Due to the weakness of this water line with respect to the He triplet, we fitted its profile from the telluric model instead of the raw spectra.

In a second phase, we considered the effects of stellar activity on the He triplet. We first studied the SiI line at $10830 \ \rm \AA$, whose properties were extracted with our custom model. Studying this line is crucial as its known stability and lack of nearby telluric lines suggest that variability in its best-fit parameters is associated with stellar activity.  

The CCF, commonly used to study the overall spectrum or evaluate the quality of the normalization procedure, is also a proxy for stellar activity. The CCF is a mathematical technique that cross-correlates the observed spectrum to a template composed of the most prominent spectral lines, giving as output an average profile of all considered spectral lines \citep[][]{Baranne-1996, Pepe-2002a}. CCF properties, such as the FWHM and bisector span, can be used to evaluate the impact of magnetic stellar activity induced by the presence of active regions on the solar surface \citep[e.g.,][]{Queloz-2001, Desort2007, Dumusque2014} or granulation and supergranulation \citep[e.g.,][]{Dumusque2016, Cegla:2019aa}. However, it is important to acknowledge that factors other than stellar activity can impact the CCF. Instrumental effects, such as variations in the point-spread function or detector nonlinearity, alongside variations in atmospheric conditions, can induce changes in CCF properties like the FWHM and contrast. 

We evaluate the impact of each mechanism by performing a correlation study. This is done by plotting the time series of each helium triplet best-fit parameter against all other quantities mentioned above. We note that our correlation studies include the HeI line's depth, width, and EW, rather than relying solely on the EW. The EW is the spectral line's area, and, therefore, variations in this quantity reflect variations in the line depth and width. These can vary in compensating ways, such as the line getting both deeper and narrower. Thus, relying only on the EW can lead to misleading correlation analyses, as a lack of correlation with the EW might mask significant variations in the line depth or width. For example, in Table \href{https://zenodo.org/records/14674319}{E.1}, correlations with the HeI depth, but not with the EW, indicate compensating variation in depth and width.

We computed Pearson and Spearman correlation coefficients for each comparison, which can capture linear and nonlinear monotonic correlations, respectively. We did so using the spearmanr and corrcoeff functions from Scipy and Numpy, respectively. Example correlation plots are shown in Appendix \href{https://zenodo.org/records/14674319}{E}. The implications of these correlations are discussed in Sect. \ref{sec:discussion_short_variability}.

\begin{figure*}
    \centering
    \includegraphics[width = \textwidth]{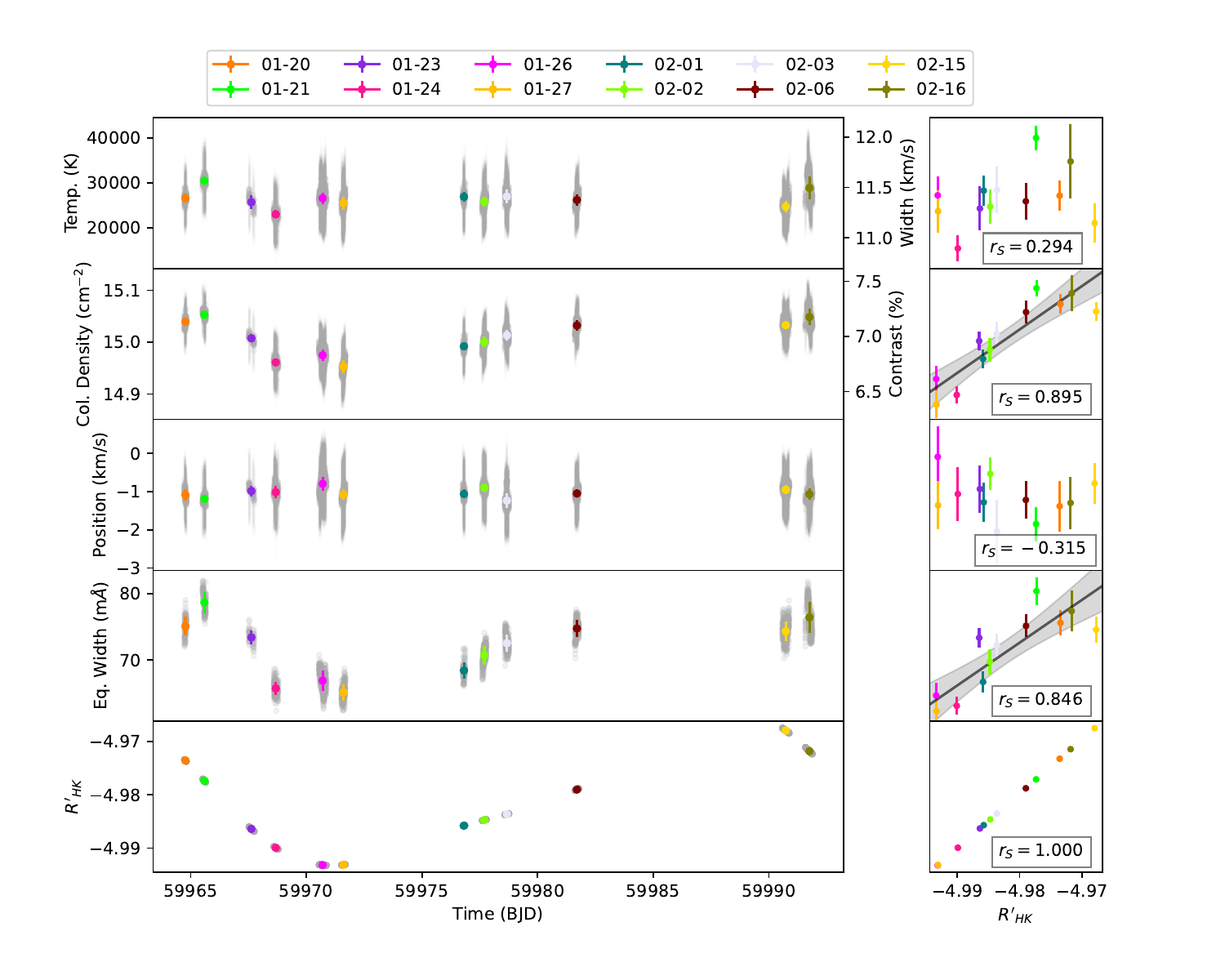}
    \caption{Time series of the He triplet properties and calcium activity index ($R'_{HK}$) over all days of observation. Left column: From top to bottom are shown the width, depth, position, and EW of the He triplet line, along with the $R'_{HK}$ activity index. Data binned over each day of observation are displayed in various colors, while unbinned data are shown in gray. Sinusoidal variations in the $R'_{HK}$ index highlight the presence of active regions during our observations. This is further corroborated by the Solar Dynamics Observatory (SDO) images shown in Figs. \ref{fig:SDO_26} and \ref{fig:SDO_27}. Right column: He triplet properties plotted against the $R'_{HK}$ index, considering only the binned data. Each subplot's legend displays the corresponding Spearman correlation coefficient, $r_S$. The black lines denote the linear fit to the correlation studies with the highest coefficients, while the shaded regions show the corresponding 95\% confidence intervals. We discuss these proportionality relationships in Sect. \ref{sec:discussion_long_variability}.
    We provide axes for the width in km/s and depth in \% to facilitate the understanding of the analogy between temperature and width, and column density and depth.}
    \label{fig:Long_timescales_+_RHK}
\end{figure*}

\subsubsection{Long timescales}\label{sec:long_timescale}
After measuring short-term variations, our focus shifted to investigating the He triplet's behavior over longer timescales. We consolidated individual time series for each He triplet property over the 12 days of observations into a unified time series. We performed correlation studies between these unified time series and the $R'_{HK}$ activity index. This index is associated with variations in the cores of the Ca II H and K lines that are closely linked to changes in the stellar magnetic field \citep[e.g.,][]{Wilson1968,Borgniet2015,Milbourne2021,Cretignier2024}. As these lines are in the visible wavelength range, the index was derived from HARPS solar spectra. However, since HARPS and NIRPS observations were conducted at different times (from 1.8 to 7.3 hours apart), the $R'_{HK}$ values were first extrapolated onto the corresponding NIRPS observation windows (see Table \ref{tab:Observation_table}) before performing our analysis. To achieve this, we performed a Fourier expansion of the activity index, keeping up to the third term. The fundamental frequency used in the expansion was obtained from a periodogram analysis and corresponded to the synodic solar rotation period. The results of our correlation studies are shown in Fig. \ref{fig:Long_timescales_+_RHK}.

\subsection{Injection-recovery tests}\label{sec:planet_signal}
In the upcoming sections, we define the methods used to generate a planetary signal, introduce it into observed solar spectra, and subsequently retrieve it. We aimed to evaluate how variations in the solar He triplet over various timescales impacted a simulated planetary signal. 

\subsubsection{Signal generation}\label{sec:planet_signal_generation}
By making similar simplifying assumptions (i.e., ignoring the exoplanet orbital geometry, and the atmosphere's emission and three-dimensional structure), we adapted the model detailed in Sect. \ref{sec:model_fit_Si_He} to simulate planetary signals. As mentioned previously, this model applies an absorption signature to a base profile. In contrast to our previous implementation, where the stellar continuum served as the base profile, we used the processed solar spectra as the base profile. In doing so, the introduced absorption signature acted as a mock planetary signal. While the planetary signal should ideally be introduced in the "true" solar spectra before performing instrumental convolution, we believe introducing it afterward yields comparable results especially since our model is a first-order approximation. Broadening induced by the exoplanet's rotation was ignored since it was negligible compared to the other sources of broadening. 

Unlike the model defined in Sect. \ref{sec:model_fit_Si_He}, the planetary absorption signature applied to the processed solar spectra only included contributions from the planetary He triplet. Gaussian profiles were used to model the three He triplet transitions. These profiles shared the parameters controlling their temperature ($T$) and column density ($n$). The $T$ and $n$ values of the simulated planetary signal were fixed to typical exoplanet thermosphere values, as discussed in Sect. \ref{sec:discussion_atm_retrieval}. The radial velocity shift, $rv$, was set to zero for simplicity.
\begin{figure}
    \centering
    \includegraphics[width=\columnwidth]{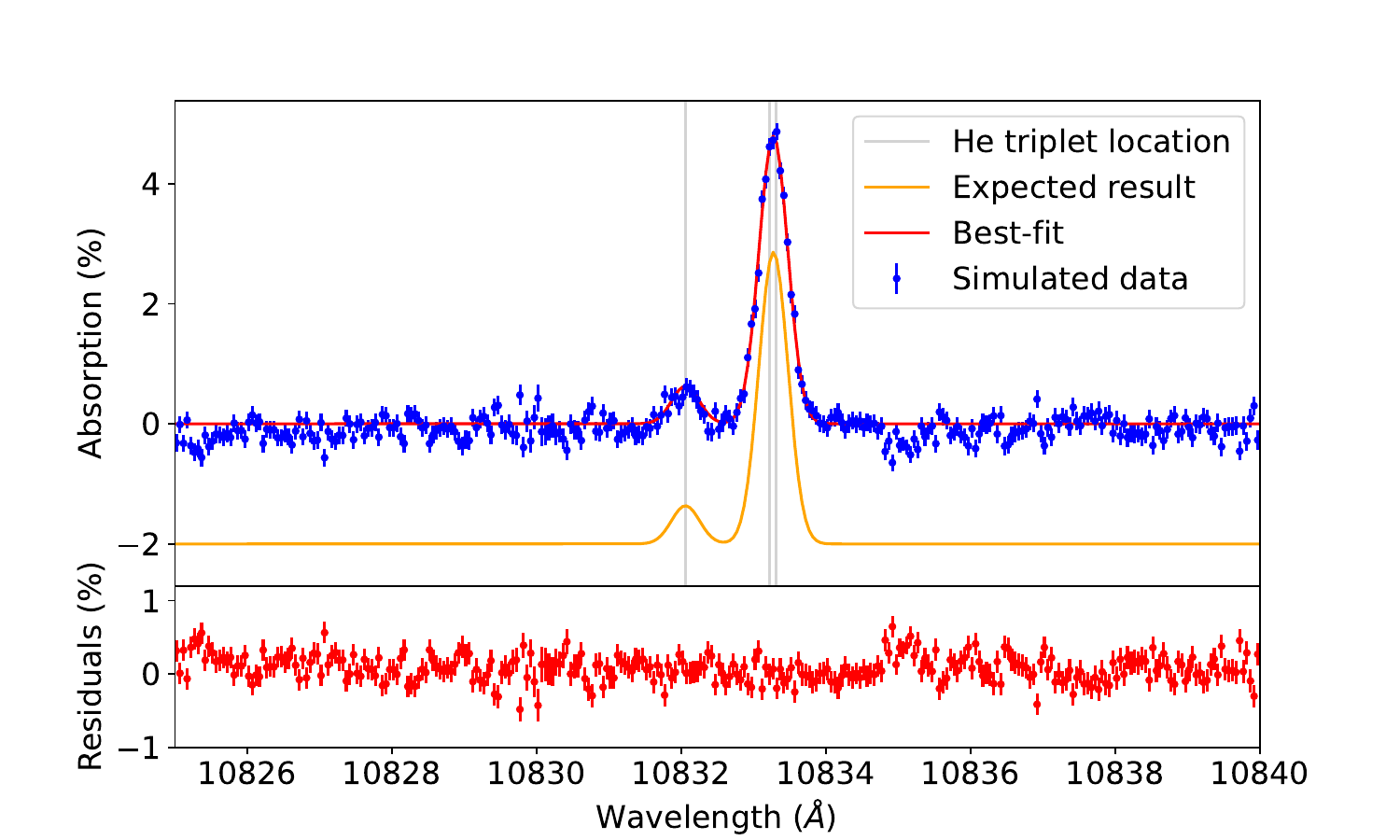}
    \caption{Mock absorption spectrum generated with an in-transit spectrum from the observation day January 26, 2023. Top panel: The blue points represent the simulated absorption spectrum and the corresponding best-fit is shown as a red line. The planetary signal, used as input for the simulation, is shown as a yellow line and is shifted vertically for display purposes. The gray vertical lines indicate the position of the He triplet. A visual inspection reveals a strong agreement between the best-fit line and the expected results. Bottom panel: Residuals for the best-fit line are shown in red.}
    \label{fig:mock_abs_spectrum}
\end{figure}

\subsubsection{Signal introduction and retrieval}\label{sec:imprint}
Having established a method to generate a mock planetary signal, we now describe how the signal is incorporated into the solar spectra, and its properties are subsequently retrieved. 
We split the array of processed solar spectra into two sets, in-transit and out-of-transit, by defining a transit duration during which the simulated planet passes in front of the Sun. For observations taken over the time interval [$t$, $t + \Delta t$], we defined [$t + \frac{1}{4} \Delta t$, $t + \frac{3}{4} \Delta t$] as in-transit and [$t$, $t + \frac{1}{4} \Delta t$]$\cup$[$t + \frac{3}{4} \Delta t$, $t + \Delta t$] as out-of-transit. In our case, $t$ denotes the timestamp at the start of a given day, and $\Delta t$ is the total duration of observations for that day. To further simplify our approach, we ignored the drop in flux during transit as it is locally achromatic, and assumed the planetary signal to be static in time. Additionally, we ignored planet-occulted line distortions as their impact on derived planetary atmosphere parameters is small, thanks to the Sun's low $v\mathrm{sin}(i)$ \citep{Dethier2023}.

\begin{figure*}
    \centering
    \includegraphics[width = \textwidth]{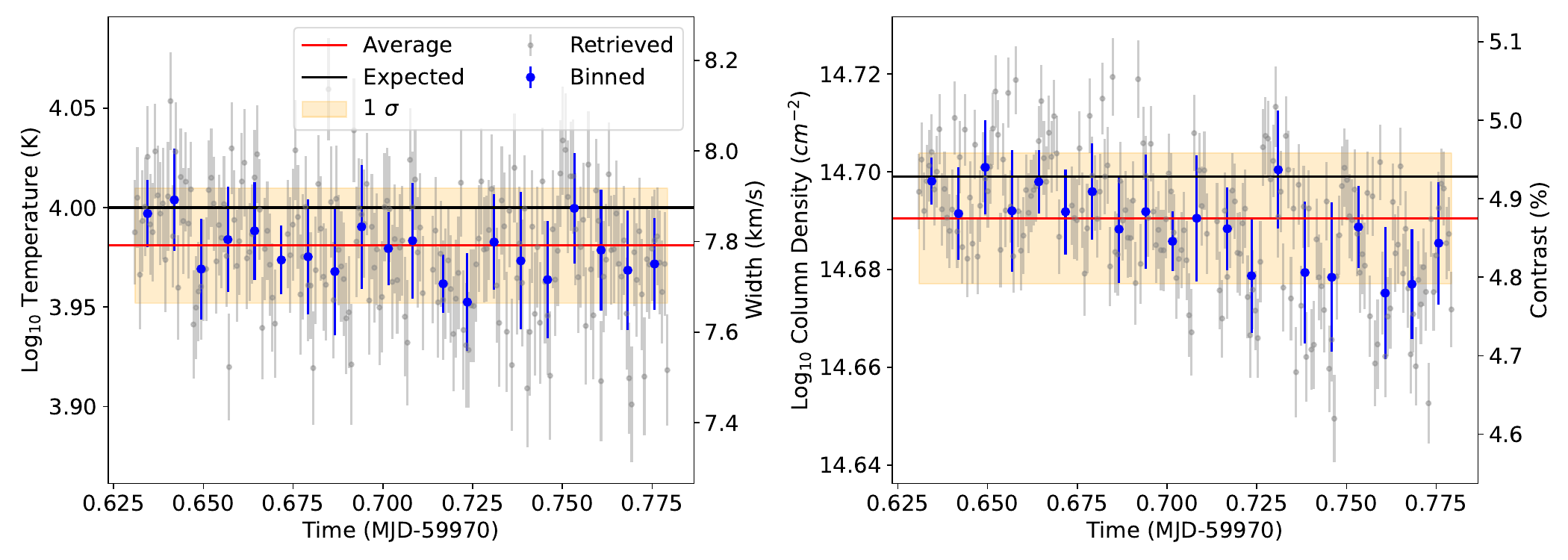}
    \caption{Time series of atmospheric temperature and He column density for the observation day January 26, 2023. Blue and gray points represent the 20x binned and unbinned best-fit planetary parameters obtained from fitting the planetary absorption spectra as described in Sect. \ref{sec:imprint}. The red line marks the weighted average of each retrieved parameter and the orange band represents the 1 $\sigma$ region around this average. The black line shows the input values used in the simulation. We provide axes for the width in km/s and depth in \% to facilitate the understanding of the analogy between temperature and width, and column density and depth.}
    \label{fig:planet_simulation}
\end{figure*}

After introducing the planetary signature into the in-transit set, we built master in-transit and out-of-transit spectra. Each master spectrum was created by taking the weighted average of the processed solar spectra in the associated set, with weights corresponding to the inverse squared flux errors. A planetary absorption spectrum, $A$, was extracted for each in-transit spectrum, $F_{in}$, with the following formula: 
\begin{equation}\label{eq:absorption_spctr}
    A = \frac{\bar{F}_{out} - F_{in}}{\bar{F}_{out}},
\end{equation}
with $\bar{F}_{out}$ the master out-of-transit spectrum. 
To retrieve the simulated planet's He triplet properties, which may have been affected by the solar He triplet, we fitted each planetary absorption spectrum with the model described in Sect. \ref{sec:planet_signal_generation}, keeping $T$ and $n$ free. Since $n$ and $T$ can reach large values, we fitted their logarithms instead. An example simulated absorption spectrum is shown in Fig. \ref{fig:mock_abs_spectrum}. We plot the time series of the resulting best-fit parameters in Figs. \ref{fig:planet_simulation} and \ref{fig:big_planet_simulation}.
To validate the results, we also retrieved the best-fit parameters for the master planetary absorption spectrum, obtained by applying Eq. \ref{eq:absorption_spctr} to the master in-transit spectrum. Finding that both master best-fit parameters were within $<0.01\sigma$ of the expected results, we did not include them in our figures.
By comparing the weighted average of the retrieved parameters to the true values used as input for the simulation, we evaluated the impact of the solar He triplet variability on retrieved planetary parameters.

We performed these injection-recovery tests both on individual days and across the entire time series of observations. This approach allowed us to study the influence of short- and long-term He triplet variations on the retrieved exoplanet properties. For the full time series' injection-recovery test, the master spectra were constructed on a per-day basis. Specifically, we treated each day as an independent transit event, using its corresponding out-of-transit and in-transit spectra to generate the associated absorption spectra.
We present and discuss the results of our various injection-recovery tests in Sect. \ref{sec:discussion_atm_retrieval}.

\section{Results and discussion}\label{sec:results_and_discussion}
We showcase our results with observations from January 26, 2023 and January 27, 2023. We select these days based on the abundance and quality of measurements, as well as the significant differences that we see in their correlation analyses that highlight the variable nature of solar He triplet measurements. We present images of the Sun's surface for those days, captured at 1700 $\rm \AA$ with the Atmospheric Imaging Assembly (AIA) of the SDO \citep{Lemen2012}, in Figs. \ref{fig:SDO_26} and \ref{fig:SDO_27}.

\subsection{Short-term variability}\label{sec:discussion_short_variability}
As shown with the residual map presented in Fig. \ref{fig:master_and_residuals}, we observe notable variability in the solar He triplet. Further support for this variability is obtained by fitting the solar He triplet and examining the time series of the fit parameters, as shown in Fig. \ref{fig:best_fit_He_param}. In particular, we observe variations on two distinct timescales: rapid oscillations spanning several minutes and a longer trend over the entire day of observation. We witness similar variations for all days of observations. 

In our analysis of the variability (see Sect. \ref{sec:diagnostics}), we explore two potential sources: stellar activity and telluric contamination. We obtain polarizing results from correlation studies between the solar He triplet and the selected stellar activity probes. 
While certain days exhibit strong correlations ($r_S > 0.5$) with multiple CCF properties, such as the FWHM, bisector span, and RV, we consistently find no significant correlations between the properties of the SiI line and the He triplet. The latter is expected, as these lines form in different regions of the solar atmosphere and the SiI line is generally stable. Interestingly, preliminary analyses revealed strong He-Si correlations, particularly in their EWs, that were highly sensitive to the normalization method used. In Appendix \ref{sec:He-Si_Correlation_Normalization} we explore this dependence and its possible origins, and discuss our motivation for adopting RASSINE normalization in our fiducial analysis. The differences in correlations with stellar activity probes across different days, or lack thereof, are exemplified in Figs. \href{https://zenodo.org/records/14674319}{E.1}, \href{https://zenodo.org/records/14674319}{E.2}, \href{https://zenodo.org/records/14674319}{E.3}, and \href{https://zenodo.org/records/14674319}{E.4} (CCF), and Figs. \href{https://zenodo.org/records/14674319}{E.5}, \href{https://zenodo.org/records/14674319}{E.6}, \href{https://zenodo.org/records/14674319}{E.7}, and \href{https://zenodo.org/records/14674319}{E.8} (Si).

Additionally, some days show strong correlations between He triplet properties and multiple telluric contamination tracers (e.g., airmass, telluric water line properties, O$_2$, and CO$_2$ ICD). Remarkably, these are the same days that exhibit strong correlations with the CCF properties (see Table \href{https://zenodo.org/records/14674319}{E.1}). This suggests that improper telluric correction may contribute to the observed He triplet variability. Indeed, telluric contamination, traced with the airmass and ICDs, can impact CCF properties such as position and shape. Furthermore, low-amplitude telluric absorption lines, such as the water line near the He triplet, are challenging to account for with the NIRPS telluric correction pipeline. This limitation can result in the He triplet varying similarly to the telluric water line properties. These results are exemplified in Fig. \href{https://zenodo.org/records/14674319}{E.9} (airmass), Figs. \href{https://zenodo.org/records/14674319}{E.10}, \href{https://zenodo.org/records/14674319}{E.11}, and \href{https://zenodo.org/records/14674319}{E.12} (ICDs), and Figs. \href{https://zenodo.org/records/14674319}{E.13}, \href{https://zenodo.org/records/14674319}{E.14}, \href{https://zenodo.org/records/14674319}{E.15}, and \href{https://zenodo.org/records/14674319}{E.16} (water line). We further emphasize the link between CCF properties and telluric contamination in Table \href{https://zenodo.org/records/14674319}{E.2}. Notably, the CCF properties most strongly correlated with the He triplet (i.e., CCF RV and FWHM) also show the strongest correlations with the telluric contamination proxy. Moreover, the lack of significant correlations on January 26, 2023 contrasted with their presence on January 27, 2023 mirrors the pattern seen in Table \href{https://zenodo.org/records/14674319}{E.1}. This provides additional evidence that the observed correlations (or lack thereof) between He triplet properties and CCF properties arise from telluric contamination.

Despite identifying telluric contamination as a possible source for the solar He triplet variability, our analysis examined each mechanism independently. Investigating the interactions between these processes through comprehensive modeling would provide a more reliable explanation for the observed variations.

Although our analysis focuses on the solar He triplet variations across an entire day of observation, short-term oscillations, lasting from several minutes to an hour, are also present (see Figure \ref{fig:best_fit_He_param}). These rapid oscillations were not studied in detail due to their negligible impact on atmospheric retrievals. Indeed, they are typically smoothed out when combining multiple transit observations. They may be of stellar or instrumental origin, warranting further investigation. 

\subsection{Long-term variability}\label{sec:discussion_long_variability}
As illustrated in Fig. \ref{fig:Long_timescales_+_RHK}, the solar He triplet properties also exhibit variations over the entire duration of our study. Both the depth and EW of the He triplet present significant correlations with the $R'_{HK}$ index. Due to the pertinence of these properties in atmospheric studies (e.g., depth provides measurements of the exospheric helium content), we provide linear relationships linking the $R'_{HK}$ index to stellar He triplet depth and EW so that future studies can evaluate or predict the impact of stellar activity on observed stellar He triplet variations.
\begin{equation}\label{eq:long_timescale_equation}
\begin{aligned}
    \mathrm{EW \ (\AA)} &= (3.8 \pm 0.8 \ \mathrm{\AA}) \cdot R'_{HK} + (34 \pm 4 \ \mathrm{\AA})\\
    \mathrm{log_{10}}n &= (0.5 \pm 0.1) \cdot R'_{HK} + (2.5 \pm 0.6)
\end{aligned}
\end{equation}
Since $R'_{HK}$ is sensitive to variations in magnetic field, this correlation could be attributed to rotational modulation due to magnetic regions on the Sun's surface. These active regions affect the flux at the photosphere and the shape of spectral lines \citep[e.g.,][]{Basri1989,Shcherbakov1996,Cretignier2021,Cretignier2023}, thus leading to changes in the line depth and EW. These findings align with \cite{Guilluy2020}, confirming the He triplet's sensitivity to variations in magnetic field. As illustrated in Figs. \ref{fig:SDO_26} and \ref{fig:SDO_27}, the witnessed rotational modulation can be attributed to the presence of faculae on the solar surface.

Our long-timescale results diverge from those presented by \cite{Krolikowski2023}. Although our EW's mean absolute deviation ($\mathrm{MAD}_{EW} = 2.35 \ \mathrm{m\AA}$) agrees with their power-law fit in Fig. 8, we find opposing results when investigating the origin of the EW variations. Indeed, while we find a strong correlation between the He triplet EW and the solar rotation (as traced by the $R'_{HK}$ index), a majority of stars in their sample do not present a similar correlation. Their sole outlier, V1298 Tau, shows a strong correlation with stellar rotation over a brief time range that they attribute to an intense flaring or mass loss event.
Such discrepancies in our results may arise from differences in the stellar populations considered. The Sun, an old and relatively quiet star, could have its long-timescale He triplet variability attributed to star spots and faculae, whereas He triplet variability in young stars may be dominated by other stellar activity processes unaffected by rotational modulation.
In light of these results, we recommend that future studies complement near-IR spectroscopic observations with concurrent photometric monitoring and optical spectroscopic observations of known activity tracers. This approach would help identify and characterize potential stellar He triplet variability. As emphasized by \cite{Krolikowski2023}, this is particularly important for observations of young stars.

\subsection{Impact on planetary atmosphere retrieval}\label{sec:discussion_atm_retrieval}
We investigate the impact of stellar He triplet variability on retrieved planetary parameters through injection-recovery tests. The results, presented in Table \ref{tab:simulation_results_table}, demonstrate that the expected parameter values consistently fall within the 1$\sigma$ region for all the days of observation, where $\sigma$ is the weighted standard deviation of the retrieved parameters. An example showcasing these results is shown in Fig. \ref{fig:planet_simulation}. The baseline case used to study the response over all the days of observation is $T = 10,000 \ \rm K$ and $n_{\rm He} = 5\times 10^{14} \ cm^{-2}$, inducing an absorption signature of $\sim 5\%$. These values reflect conditions in the upper atmospheric layers of hot giant planets, which are the main targets of atmospheric escape studies due to their significant transmission spectroscopy signatures. These layers can reach temperatures ranging from 5,000 to 20,000 K \citep[e.g.,][]{Guilluy2020, Lampon2023} and, being primarily composed of light elements such as H and He, can induce absorption signatures of several percent \citep[e.g.,][]{Allart2018, Allart2019}, making the choice of input conditions realistic. 

From these observations, and the values presented in Table \ref{tab:simulation_results_table}, we conclude that the solar He triplet variability, present on both ultra-short timescales ($\sim$ 10 min) and over the entire day of observation (as highlighted by the binning in Fig. \ref{fig:planet_simulation}), does not have a significant impact on the retrieved planetary parameters. Thus, assuming the stellar He triplet is stable throughout the transit duration remains a valid approximation. Furthermore, our model, although a first-order approximation reliant on basic assumptions, successfully demonstrates the minimal impact of solar He triplet variability on the retrieved planetary parameters. This implies that introducing additional complexity to the model would lead to the same conclusion.

\begin{figure*}
    \centering
    \includegraphics[width = \textwidth]{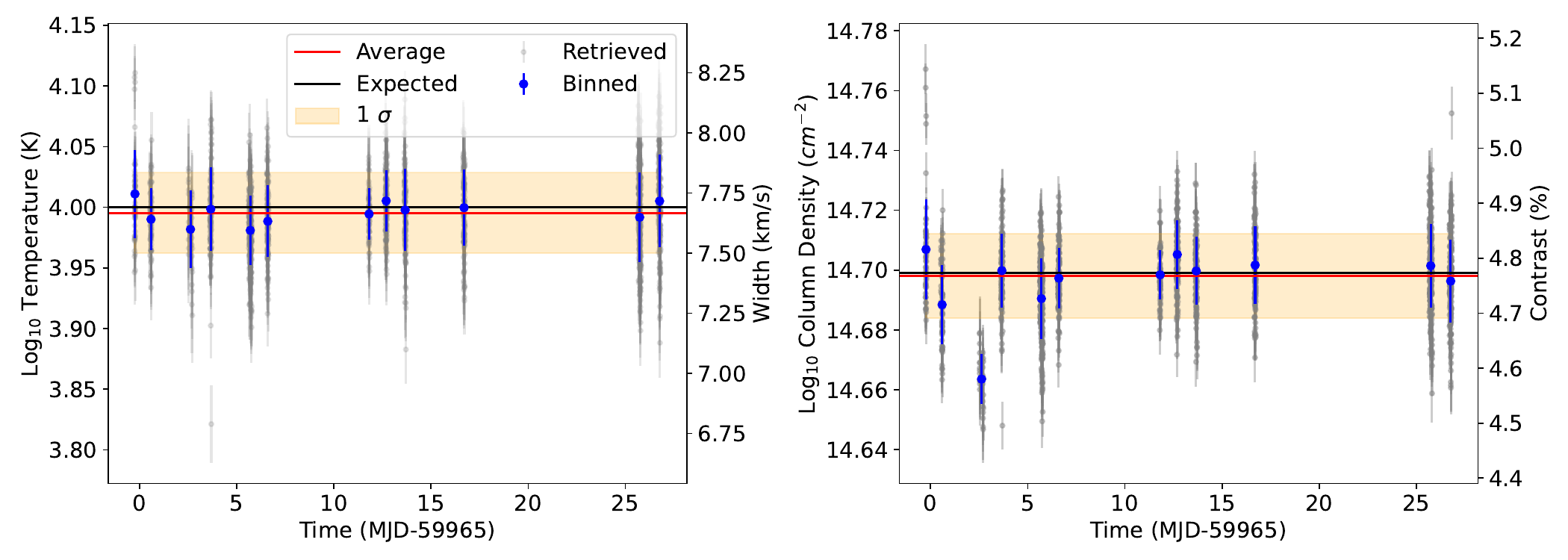}
    \caption{Time series of atmospheric temperature and He column density over all days of observation. Blue and gray points represent the daily binned and unbinned best-fit planetary parameters obtained from fitting the planetary absorption spectra as described in Sect. \ref{sec:imprint}. The lines, band, and axes are the same as in Fig. \ref{fig:planet_simulation}.}
    \label{fig:big_planet_simulation}
\end{figure*}

Our injection-recovery analysis of long-term He triplet variability, presented in Fig. \ref{fig:big_planet_simulation} and the last row of Table \ref{tab:simulation_results_table}, yields similar results: the expected parameter values remain within the corresponding 1$\sigma$ region.
This result suggests that variations in the He triplet over days to months, due to stellar activity, do not significantly affect the retrieval of exoplanet properties. 

We stress that our results are only valid for our analysis of a $5\%$ planetary signature in the presence of solar variability. Further analyses would have to be conducted to evaluate the impact of stellar He triplet variability when more or less active stars and stronger or weaker planetary signatures are considered.

\section{Conclusion}\label{sec:conclusions}
Our study revealed the presence of variability in the solar He triplet, a spectral line recently used in the study of atmospheric escape. We demonstrate that the He triplet exhibits variations across different timescales under the influence of various processes. We identify improper telluric correction as the potential source of short-term He triplet variability and stellar activity as the definitive source of long-term He triplet variability. We elected not to investigate the impact of instrumental noise (e.g., modal noise) on our results as it is hard to quantify, and its impact is likely negligible. Despite having identified sources for the He triplet variability, we recommend that additional studies be performed accounting for correlated noise and the combined effects of stellar activity and telluric contamination, for example using the ANTARESS workflow \citep{Bourrier2024} to reduce and analyze the helium spectra. Further studies could also refine the long-timescale relations that we established between the $R'_{HK}$ index and the He triplet depth and EW (see Eq. \ref{eq:long_timescale_equation}).

Importantly, our findings indicate that variations in the solar He triplet, over the timescale of a single transit or across multiple transits, have minimal impact on the retrieved planetary parameters crucial to the study of atmospheric escape. Moreover, given our use of a first-order model, we expect that more complex modeling approaches would reach similar conclusions. While we definitively observe strong variations induced by stellar activity on the timescale of solar rotation, as seen in \cite{Guilluy2020}, our analysis demonstrates that these variations are well captured in the master-out spectrum and would therefore not impact studies focusing on the variability of exoplanet atmospheres through time.

These results are significant as the He triplet is gaining in popularity and availability for atmospheric studies, especially with the arrival of new instruments and telescopes, like NIRPS and NIGHT, that specifically target the IR range to measure planetary He signals and characterize stellar He variability.
However, we note that the conclusions drawn here are only valid for old, quiet Sun-like stars. It is well-known that stellar activity varies significantly across different spectral types and stellar ages. As a result, the He triplet variability would need to be reevaluated when considering M-dwarfs, for example.
\newline
\newline

\section*{Data availability}\label{sec:data_availability}
The plots and tables illustrating the correlation studies between the He triplet properties, stellar activity tracers, and telluric contamination tracers, referred to as Appendix E, can be found online at \url{https://zenodo.org/records/14674319}.

\begin{acknowledgements}
SJM  acknowledges support from the Massachusetts Institute of Technology through the Praecis Presidential Fellowship. XD  acknowledges the support from the European Research Council (ERC) under the European Union’s Horizon 2020 research and innovation programme (grant agreement SCORE No 851555) and from the Swiss National Science Foundation under the grant SPECTRE (No 200021\_215200). This work has been carried out within the framework of the NCCR PlanetS supported by the Swiss National Science Foundation under grants 51NF40\_182901 and 51NF40\_205606. This project has received funding from the European Research Council (ERC) under the European Union's Horizon 2020 research and innovation programme (project {\sc Spice Dune}, grant agreement No 947634). This material reflects only the authors' views and the Commission is not liable for any use that may be made of the information contained therein. MCr acknowledges the Swiss National Science Foundation funding under grant P500PT 211024. WD \& NCS  acknowledge the support from FCT - Funda\c{c}\~ao para a Ci\^encia e a Tecnologia through national funds by these grants: UIDB/04434/2020, UIDP/04434/2020. Co-funded by the European Union (ERC, FIERCE, 101052347). Views and opinions expressed are however those of the author(s) only and do not necessarily reflect those of the European Union or the European Research Council. Neither the European Union nor the granting authority can be held responsible for them. RA  acknowledges the Swiss National Science Foundation (SNSF) support under the Post-Doc Mobility grant P500PT\_222212 and the support of the Institut Trottier de Recherche sur les Exoplan\`etes (IREx). RA, FBa \& RD  acknowledge the financial support of the FRQ-NT through the Centre de recherche en astrophysique du Qu\'ebec as well as the support from the Trottier Family Foundation and the Trottier Institute for Research on Exoplanets. FBa \& RD  acknowledge support from Canada Foundation for Innovation (CFI) program, the Universit\'e de Montr\'eal and Universit\'e Laval, the Canada Economic Development (CED) program, and the Ministere of Economy, Innovation and Energy (MEIE). AC acknowledges funding from the French ANR under contract number ANR\-18\-CE31\-0019 (SPlaSH), and the French National Research Agency in the framework of the Investissements d'Avenir program (ANR-15-IDEX-02), through the funding of the ``Origin of Life" project of the Grenoble-Alpes University. Our solar data analysis made use of the following packages:
Astropy\footnote{\url{https://www.astropy.org}}\citep[v5.3.4,][]{astropy_2013, astropy_2018, astropy_2022}, Matplotlib\footnote{\url{https://matplotlib.org}}\citep[v3.7.2,][]{Hunter2007}, Numpy\footnote{\url{https://numpy.org}}\citep[v1.26.1,][]{Harris2020}, 
Pandas\footnote{\url{https://pandas.pydata.org}}\citep[v1.5.3,][]{McKinney2010}, Scipy\footnote{\url{https://scipy.org}}\citep[v1.11.3,][]{Virtanen2020}
\end{acknowledgements}

\bibliographystyle{aa}
\bibliography{He_bibliography}

\begin{appendix}
\onecolumn
\section{HELIOS observations}

\begin{table}[h!]
    \caption{\label{tab:Observation_table}HELIOS observations used in this work.}
    \centering
    \begin{tabular}{c|c|c|c|c|c}
    \hline
        Observation date & Start time (UTC) & End Time (UTC) & Exp. numbers & Airmass range & S/N\\
    \hline
        January 20, 2023 & 17:31 & 19:26 & 113 & 1.03 - 1.23 & 664\\
        January 21, 2023 & 12:00 & 15:34 & 155 & 1.07 - 2.48 & 662\\
        January 23, 2023 & 12:00 & 17:47 & 99 & 1.02 - 2.52 & 656\\
        January 24, 2023 & 14:21 & 18:41 & 202 & 1.02 - 1.24 & 665\\
        January 26, 2023 & 13:20 & 21:12 & 462 & 1.02 - 1.95 & 660\\
        January 27, 2023 & 12:19 & 16:22 & 239 & 1.03 - 2.21 & 655\\
        February 01, 2023 & 18:37 & 20:35 & 117 & 1.12 - 1.61 & 654\\
        February 02, 2023 & 14:42 & 18:45 & 239 & 1.02 - 1.19 & 657\\
        February 03, 2023 & 14:10 & 18:10 & 236 & 1.03 - 1.31 & 654\\
        February 06, 2023 & 14:58 & 18:48 & 226 & 1.03 - 1.16 & 653\\
        February 15, 2023 & 14:15 & 22:45 & 500 & 1.04 - 6.81 & 648\\
        February 16, 2023 & 14:08 & 21:07 & 331 & 1.05 - 2.03 & 648\\
    \end{tabular}
    \tablefoot{All observations were obtained with an exposure time of 50.1597 s. We note the important air mass range for numerous days that is taken into account during data reduction (see Sect. \ref{sec:outlier_clipping}). In the last column, we report the average S/N around the He triplet feature.}
\end{table}

\section{He-Si EW correlation}\label{sec:He-Si_Correlation_Normalization}
The properties of the solar He triplet, formed in the upper chromosphere \citep{Goldberg1939, Zirin1975}, and the SiI line, formed in the upper photosphere \citep{BardCarlsson2008}, are not expected to exhibit strong correlations. Nevertheless, our preliminary analyses reveal a significant correlation in their equivalent widths. Moreover, we find that this correlation depends on the normalization method used. This puzzling result prompted us to explore various normalization methods to determine the optimal approach for normalizing spectra and understand the cause of the observed correlation's dependence on the chosen method. 

\begin{figure}[h!]
  \centering
  \begin{minipage}{.49\textwidth}
      \centering
      \includegraphics[width=\textwidth]{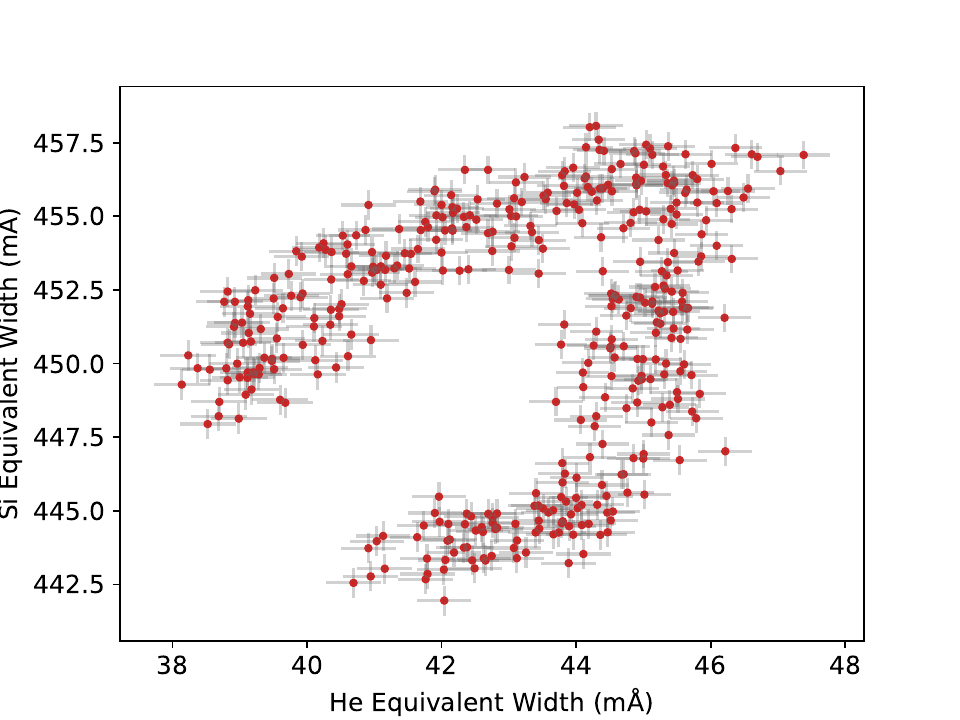}
  \end{minipage}
  \begin{minipage}{.49\textwidth}
      \centering
      \includegraphics[width=\textwidth]{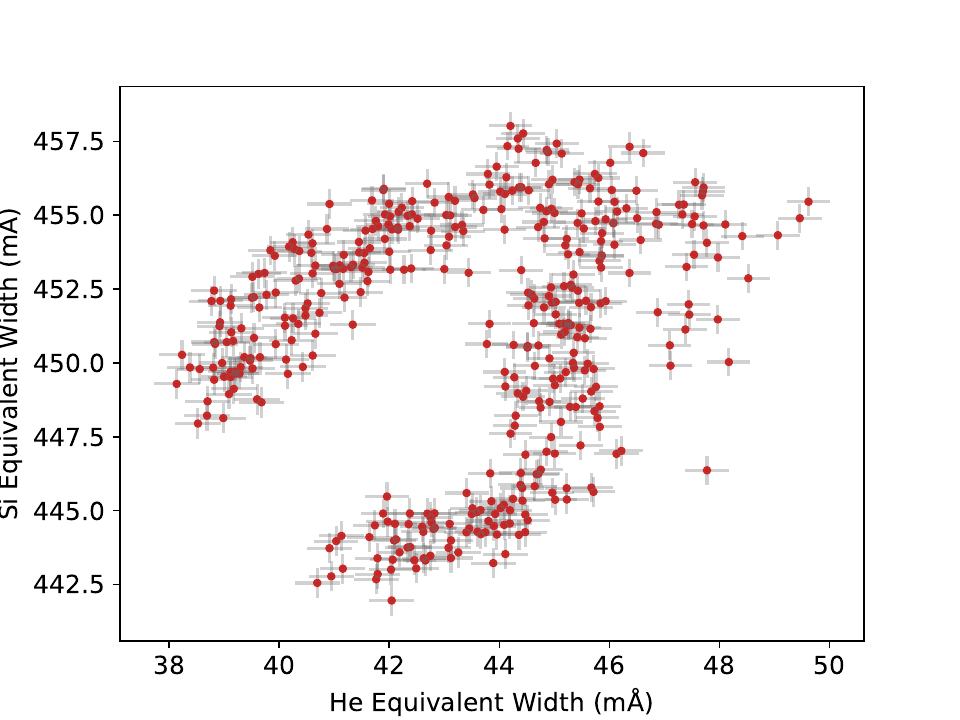}
  \end{minipage}
  \caption{\label{fig:orderpoly_SiHeEWcorr}Correlation between the He and Si EW when using a first-order (left) and second-order (right) polynomial to normalize the solar spectra. The Pearson and Spearman correlation coefficients for this plot are displayed in Table \ref{tab:normalization_result_table}. As detailed in Appendix \ref{sec:He-Si_Correlation_Normalization}, the low coefficient values do not imply the absence of correlation but rather indicate that the symmetric and complex nature of the data cannot be accurately captured by the selected correlation metrics.}
\end{figure}

In our initial approach, the stellar continuum around the He triplet of each un-normalized solar spectrum is described with an n-th order polynomial. This polynomial acts as the base profile onto which the absorption signature is applied in our model (outlined in Sect. \ref{sec:model_fit_Si_He}). As reported in Table \ref{tab:normalization_result_table}, we retrieve significant, linear or nonlinear, He-Si EW correlations when using a zeroth-order polynomial. Despite obtaining low correlations with the first- and second-order polynomials, both correlation plots display noticeable structure (see Fig. \ref{fig:orderpoly_SiHeEWcorr}). We conclude that due to the complexity, particularly the symmetry, of this structure, the Pearson and Spearman coefficients fail to accurately capture the strength of the correlation. 
The variation in correlation coefficients across different polynomial orders suggests that the observed correlation is not physical but an artifact of our normalization procedure. This prompts us to adopt a more refined approach: each solar spectrum is first normalized with the RASSINE pipeline \citep{Cretignier2020} before fitting it with our model.

As mentioned in Sect. \ref{sec:normalization}, RASSINE utilizes an alpha-shape algorithm to scan spectra and identify local maxima, also known as anchor points. A key parameter in the alpha-shape algorithm is the rolling pin radius. This user-selected parameter can have important repercussions, as too small a rolling pin can lead to anchor points being placed inside stellar lines. This can result in the normalization procedure absorbing variations intrinsic to the stellar lines, thus biasing our results. Selecting an optimal rolling pin radius is therefore a crucial step in the normalization procedure. Thankfully, the rolling pin radius can be adjusted, and its impact on the anchor points can be observed, through the RASSINE graphical user interface (GUI). Using RASSINE's GUI to determine the optimal pin radius thus ensures that anchor points are not placed inside the He triplet and SiI lines. Additionally, we perform an extra cautionary step to verify that RASSINE's normalization does not absorb any of the variations in the He triplet and SiI lines. RASSINE allows users to manually exclude wavelength ranges, preventing anchor points found in this range from contributing to the continuum profile. Combining this option with the optimal rolling pin radius we find no significant He-Si EW correlation (see Table \ref{tab:normalization_result_table}). In doing so, we prove that RASSINE's normalization method, using the optimal rolling pin radius, does not significantly absorb variations intrinsic to the He triplet and Si lines. 
The results presented in Table \ref{tab:normalization_result_table}, and Fig. \ref{fig:orderpoly_SiHeEWcorr} are for observation day January 26, 2023.

\begin{table}[h!]
    \centering
    \caption{\label{tab:normalization_result_table}Results of the He-Si EW correlation with different normalization methods.}
    \begin{tabular}{c|c|c}
         Method used & $r_P$ & $r_S$\\
         \hline
         Polynomial, n=0 & $0.904$ & $0.846$\\
         Polynomial, n=1 & $0.156$ & $0.224$\\
         Polynomial, n=2 & $0.168$ & $0.221$\\
         \hline
         RASSINE, small radius & $-0.504$ & $-0.520$\\
         RASSINE, optimal radius + range exclusion & $0.286$ & $0.240$\\
    \end{tabular}
\end{table}

\section{Planetary signal simulation}\label{sec:appendix_plan_signal_simul}
\begin{table}[h!]
    \caption{\label{tab:simulation_results_table}Results of injection-recovery test for each observation day and all observation days combined.}
    \centering
    \begin{tabular}{c|c|c|c|c|c|c}
    \hline
        Observation date & \makecell{Transit\\ duration (hrs)} & \makecell{Expected\\ T(K)} & \makecell{Retrieved\\ T(K)} & \makecell{Expected $n_{\rm He}$($m^{-2}$)} & \makecell{Retrieved $n_{\rm He}$($m^{-2}$)}& Relative difference \\
    \hline
        January 20, 2023 & 0.95 & 10000 & 10256 & 5$\times 10^{14}$ & 5.09$\times 10^{14}$ & 0.30\ $\sigma_T$ \& 0.47\ $\sigma_n$\\
        January 21, 2023 & 1.53 & 10000 & 9773 & 5$\times 10^{14}$ & 4.88$\times 10^{14}$ & 0.39\ $\sigma_T$ \& 0.80\ $\sigma_n$\\
        January 23, 2023 & 2.63 & 10000 & 9590 & 5$\times 10^{14}$ & 4.61$\times 10^{14}$ & 0.57\ $\sigma_T$ \& 4.25\ $\sigma_n$\\
        January 24, 2023 & 2.17 & 10000 & 9964 & 5$\times 10^{14}$ & 5.01$\times 10^{14}$ & 0.05\ $\sigma_T$ \& 0.07\ $\sigma_n$\\
        January 26, 2023 & 3.69 & 10000 & 9571 & 5$\times 10^{14}$ & 4.90$\times 10^{14}$ & 0.66\ $\sigma_T$ \& 0.63\ $\sigma_n$\\
        January 27, 2023 & 1.90 & 10000 & 9737 & 5$\times 10^{14}$ & 4.98$\times 10^{14}$ & 0.39\ $\sigma_T$ \& 0.16\ $\sigma_n$\\
        February 01, 2023 & 0.99 & 10000 & 9873 & 5$\times 10^{14}$ & 4.99$\times 10^{14}$ & 0.26\ $\sigma_T$ \& 0.06\ $\sigma_n$\\
        February 02, 2023 & 2.03 & 10000 & 10122 & 5$\times 10^{14}$ & 5.07$\times 10^{14}$ & 0.21\ $\sigma_T$ \& 0.54\ $\sigma_n$\\
        February 03, 2023 & 2.00 & 10000 & 9949 & 5$\times 10^{14}$ & 5.01$\times 10^{14}$ & 0.07\ $\sigma_T$ \& 0.07\ $\sigma_n$\\
        February 06, 2023 & 1.92 & 10000 & 9992 & 5$\times 10^{14}$ & 5.03$\times 10^{14}$ & 0.01\ $\sigma_T$ \& 0.21\ $\sigma_n$\\
        February 15, 2023 & 3.42 & 10000 & 9812 & 5$\times 10^{14}$ & 5.03$\times 10^{14}$ & 0.22\ $\sigma_T$ \& 0.18\ $\sigma_n$\\
        February 16, 2023 & 3.47 & 10000 & 10119 & 5$\times 10^{14}$ & 4.97$\times 10^{14}$ & 0.13\ $\sigma_T$ \& 0.19\ $\sigma_n$\\
        \hline
        Combined & N/A & 10000 & 9893 & 5$\times 10^{14}$ & 4.99$\times 10^{14}$ & 0.14\ $\sigma_T$ \& 0.07\ $\sigma_n$\\
    \end{tabular}
    \tablefoot{We specify the transit duration used to delimit in- and out-of-transit observations for each day (see Sect. \ref{sec:imprint} for its definition). The relative difference is calculated by dividing the absolute difference by the weighted standard deviation derived from each set of retrieved parameters. The last column presents the relative difference for the atmospheric temperature and He column density, in that order.}
\end{table}

\begin{figure}[h!]
\section{SDO images}
\vspace{0.3cm}
  \centering
  \begin{minipage}{.49\textwidth}
      \centering
      \includegraphics[width=\textwidth]{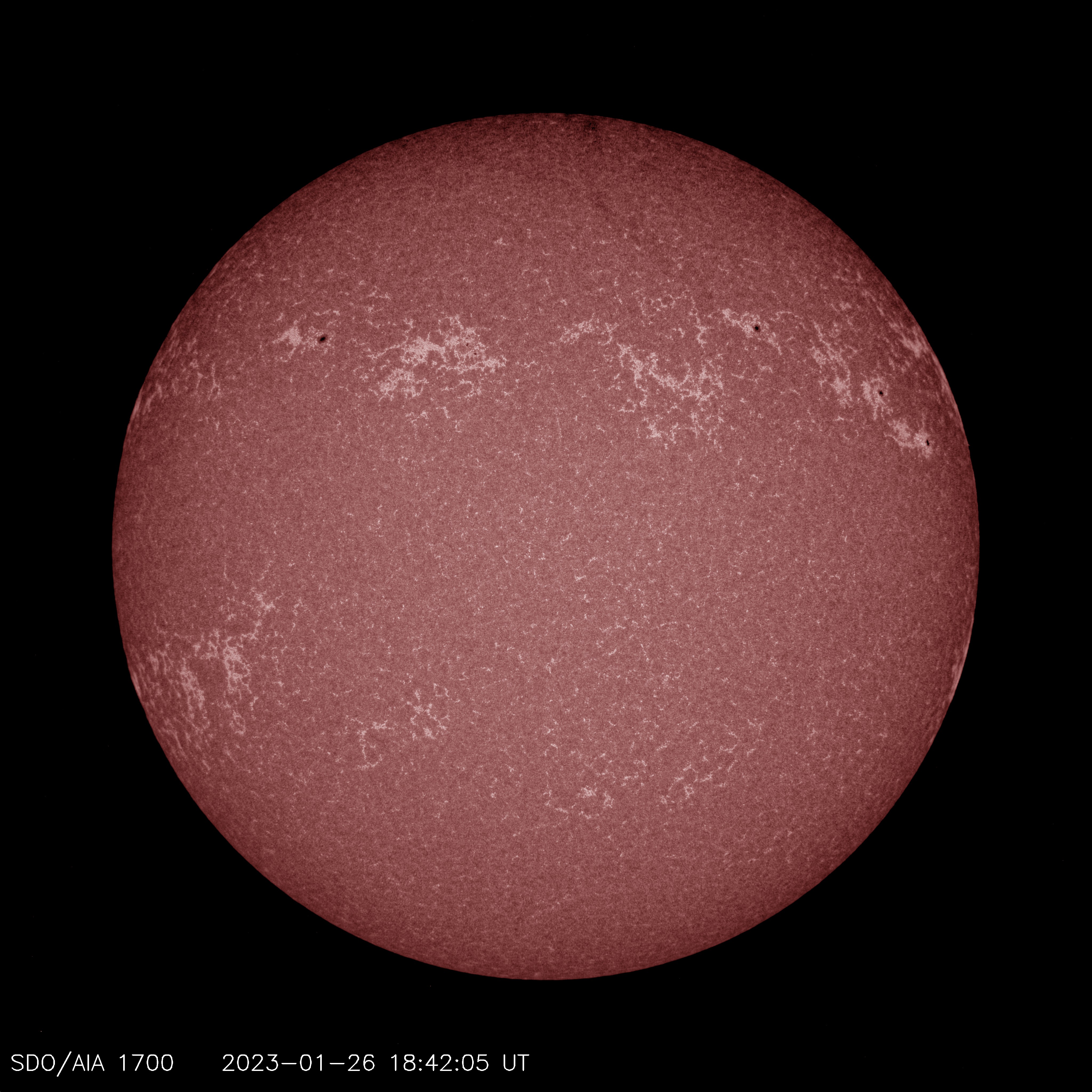}
      \captionof{figure}{\label{fig:SDO_26}January 26, 2023.}
  \end{minipage}
  \begin{minipage}{.49\textwidth}
      \centering
      \includegraphics[width=\textwidth]{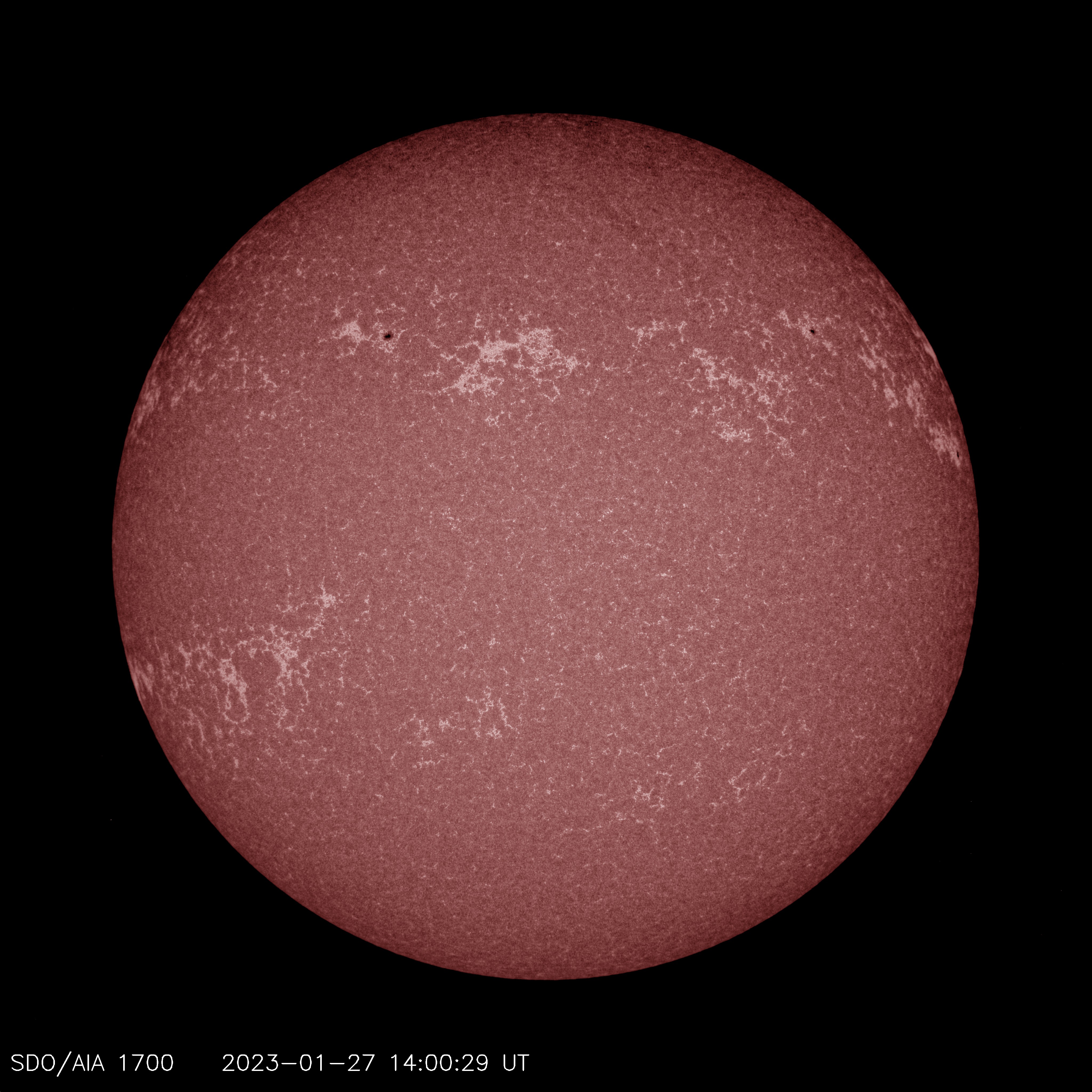}
      \captionof{figure}{\label{fig:SDO_27}January 27, 2023.}
  \end{minipage}
  \caption{SDO/AIA 1700 $\rm \AA$ images for the solar surface during the days of observation were selected to showcase our analysis. Importantly, we note the lack of spots and the presence of faculae for these two days.}
  \label{fig:SDO_images}
\end{figure}
\end{appendix}

\end{document}